\documentclass{iopjournal}

\usepackage{amsmath}
\usepackage{amssymb}
\usepackage{amsthm}
\usepackage{mathtools}
\usepackage{physics}
\usepackage[numbers,sort&compress]{natbib}
\usepackage{doi}

\newcommand{\PT}{$\mathcal{PT}$}
\renewcommand{\epsilon}{\varepsilon}
\newtheorem{theorem}{Theorem}

\renewcommand{\articletype}[1]{\vspace*{-8mm} \noindent{\scriptsize\sffamily\bfseries\MakeUppercase{#1}}\par\vspace*{8mm}}

\begin{document}

\articletype{Preprint}

\title{Non-Hermitian effective dynamics from post-selection protocols}

\author{Apostolos Siskou\hyperlink{affil:rwth}{\textcolor{black}{$^1$}}\hyperlink{author:corresponding}{\textcolor{black}{$^*$}}\orcid{0009-0001-7513-8974} and Volker Meden\hyperlink{affil:rwth}{\textcolor{black}{$^1$}}}

\affil{\hypertarget{affil:rwth}{$^1$}Institut f\"ur Theorie der Statistischen Physik, RWTH Aachen University, 52056 Aachen, Germany}

\affil{\hypertarget{author:corresponding}{$^*$}Author to whom any correspondence should be addressed.}

\email{\href{mailto:apostolos.siskou@rwth-aachen.de}{apostolos.siskou@rwth-aachen.de} and \href{mailto:meden@physik.rwth-aachen.de}{meden@physik.rwth-aachen.de}}

\keywords{non-Hermitian quantum mechanics, post-selection, generalized measurements, open quantum systems}

\begin{abstract}
Multiple seemingly distinct frameworks are commonly proposed to explain the emergence of effective non-Hermitian dynamics in quantum systems coupled to an environment. Furthermore, it remains contested how to properly calculate expectation values or whether thermal equilibrium states exist in non-Hermitian systems. In this work, we show that Feshbach projection, the Lindblad master equation, and the ancilla approach can all be understood through the Kraus map formalism, leading to non-Hermitian time evolution if an appropriate post-selection procedure is enforced. After first deriving an exact expression for the conditional system dynamics, we discuss the approximations employed in each approach and the classes of non-Hermitian Hamiltonians that can be obtained. Within this framework, we physically justify the consistency of the right-right formalism (as opposed to the biorthogonal one) for calculating expectation values. Finally, we demonstrate that the traditional notion of thermodynamic equilibrium is not applicable to general non-Hermitian systems. Consequently, stationary states are characterized as non-equilibrium steady states, which should not be conflated with system-projected thermal states of the system-environment composite.
\end{abstract}

\section{Introduction}
\label{sec:introduction}
Non-Hermitian Hamiltonians have garnered attention in recent years due to unique effects such as the non-Hermitian skin effect \cite{ZhangSkinEffectReview, SkinEffect} and a rich variety of topological phases \cite{Classification}. The consideration of such Hamiltonians is usually justified in an open-system context in one of three ways: the Lindblad master equation with post-selection for the no-jump sector, Feshbach projection in the wide-band limit, and the ancilla approach with post-selection for the spin-up sector \cite{AshidaReview, AncillaIntroduction}. While the use of non-Hermitian Hamiltonians and the necessity of a post-selection procedure for genuine non-Hermitian dynamics have been studied for decades in the quantum optics community \cite{Carmichael}, a shared framework for the physical origins of such dynamics is necessary in the quantum many-body context in light of possible future experimental realizations and their interpretation. This work unifies the three approaches through the framework of generalized measurements, showing that the occurrence of non-Hermitian dynamics is the result of the nonlinear nature of quantum mechanical measurements and conditioning on a certain measurement outcome. The resulting framework provides a rigorous, operationally grounded account of non-Hermitian dynamics, from which both their physical origin and a consistent prescription for calculating physical observables follow naturally.

As we will show, genuine non-Hermitian Hamiltonians, i.e., generators of time evolution acting on quantum states, are a manifestation of Bayesian conditional probability given that a system-environment composite survives a post-selection protocol up to a time $t$. The unnormalized states then follow the Schr\"odinger equation
\begin{equation} \label{eq:schroedinger}
i \partial_t \ket{\Psi^u(t)} = \hat{H}_{\mathrm{eff}}(t) \ket{\Psi^u(t)}
\end{equation}
with a non-Hermitian Hamiltonian $\hat{H}_{\mathrm{eff}}(t)$. The squared norm of $\ket{\Psi^u(t)}$ measures the post-selection success probability relative to its initial value, while the physical states remain normalized and follow the nonlinear equation of motion
\begin{equation} \label{eq:nonlinschroedinger} 
i \partial_t \ket{\Psi (t)} = \hat{H}_{\mathrm{eff}}(t) \ket{\Psi (t)} - \ev{\hat{H}_{\mathrm{eff}}^{a}(t)}{\Psi(t)} \ket{\Psi(t)}
\end{equation}
with $\hat{H}_{\mathrm{eff}}^{a}(t) = \frac{1}{2} [ \hat{H}_{\mathrm{eff}}(t) - \hat{H}_{\mathrm{eff}}^\dagger (t)]$ being the anti-Hermitian part of $\hat{H}_{\mathrm{eff}}(t)$ \cite{GisinNonLinear, NonLinearEquationOfMotion}.

Within this unified picture, we extract specific constraints and physical interpretations for non-Hermitian systems. 

First, we show that eigenvalues of a time-independent $\hat{H}_{\mathrm{eff}}$ with a nonzero imaginary part indicate an exponential growth or decline of the relative post-selection probability (``gain and loss''), while transient sinusoidal oscillations of the norm arise due to the non-orthogonality of the Hamiltonian's right eigenstates. We delineate the boundaries between passive and active gain, deriving theoretical limits on the maximum time for which the ancilla construction remains valid and explaining them from a physical standpoint (see also the supplemental material of Ref.~\cite{WuPTSymmetry}). 

Second, we physically justify that the right-right (RR) formalism for the calculation of expectation values is the only approach consistent with generalized measurements. This invalidates the prescription of a biorthogonal inner product to calculate expectation values of observables. The latter is often used in the literature not only to calculate spectral properties but also to evaluate physical quantities like the current or density (e.g., in Refs.~\cite{KarmakarLRCurrent, MoirePlasmonicsLR}), and results in unphysical, complex values; for an earlier discussion of this, see Ref.~\cite{PTpaper}.

Finally, we demonstrate that a Gibbs ensemble derived via the maximization of the von Neumann entropy given physically chosen constraints (e.g., the mean energy or particle number) is not stationary for generic non-Hermitian Hamiltonians. Consequently, the traditional notion of thermodynamic equilibrium is generally not applicable and stationary states are non-equilibrium steady states.

The paper is organized as follows. Section \ref{sec:generalizedmeasurements} introduces the concept of generalized measurements and presents how post-selection leads to exact, non-Hermitian dynamics. Section \ref{sec:approaches} derives the emergence of non-Hermitian Hamiltonians from the Feshbach projection, the Lindblad master equation, and the ancilla approach by identifying the corresponding post-selected trajectory and the employed approximations compared to the exact dynamics. Section \ref{sec:interpretation} addresses the physical consequences for the inner product and expectation values. Section \ref{sec:equilibrium} examines the applicability of standard thermal equilibrium in non-Hermitian systems and determines the general form of non-Hermitian steady states. Finally, Section~\ref{sec:conclusion} summarizes the main results and their implications.

\section{Generalized measurements} \label{sec:generalizedmeasurements}
In order to describe measurements on a system that becomes entangled with an environment through unitary time evolution, we employ the concept of generalized measurements, following Refs.~\cite{harochebook, nielsen}. To avoid domain issues associated with unbounded operators, we work with finite-dimensional Hilbert spaces (or suitable finite-dimensional truncations) for the effective non-Hermitian subsystems wherever necessary.

Each possible generalized measurement outcome $i$ is assigned an operator $\hat{M}_i$ acting on the system Hilbert space $\mathcal{H}_S$. The set $\{ \hat{M}_i \}$ has to fulfill the completeness relation
\begin{equation}
\label{eq:completeness}
\sum_i \hat{M}_i^{\dagger} \hat{M}_i = \hat{1}.
\end{equation}
We assume that the system is in a mixed state with density operator $\hat{\rho}_S$. The probability $p_i$ of measurement outcome $i$ and the new density operator $\hat{\rho}_{S|i}$ after measuring $i$ are then defined as
\begin{equation}
\label{eq:probability}
p_i = \Tr(\hat{M}_i^{\dagger} \hat{M}_i \hat{\rho}_S) \quad \text{and} \quad \hat{\rho}_{S|i} = \frac{\hat{M}_i \hat{\rho}_S \hat{M}_i^{\dagger}}{p_i},
\end{equation}
respectively. Equation~\eqref{eq:completeness} ensures that the probabilities add up to unity due to the normalization of the density operator. Moreover, because $\hat{M}_i^{\dagger} \hat{M}_i$ is Hermitian and positive semi-definite, $p_i \geq 0$. The set of all $\hat{M}_i^{\dagger} \hat{M}_i \eqcolon \hat{R}_i$ constitutes a positive operator-valued measure (POVM) \cite{daviesbook}. Note that a POVM does not contain the full information about the generalized measurement, since the state of the system after the measurement cannot be constructed from the $\hat{R}_i$ alone. 

Standard projective measurements known from the study of closed systems are included in the definition of Eq.~\eqref{eq:probability} when $\hat{M}_i \equiv \hat{P}_i = \hat{P}_i^2 = \hat{P}_i^{\dagger}$ are mutually orthogonal projectors, i.e., $\hat{P}_i \hat{P}_j = \delta_{ij} \hat{P}_i$. 

It has to be stressed that the concept of generalized measurements does not necessitate additional quantum mechanical postulates. We show this by tracing back its interpretation to projective measurements on two initially unentangled subsystems acted on by a unitary time evolution operator $\hat{U}(t)$. The Hilbert space of the composite is given by $\mathcal{H}_{SE} = \mathcal{H}_S \otimes \mathcal{H}_E$, where $\mathcal{H}_E$ is the Hilbert space of the environment. Assuming that the environment is initially ($t=0$) in a pure state $\ket{0_E}$ belonging to an orthonormal basis $\{ \ket{n_E} \}$ of $\mathcal{H}_E$, the time-evolved system-environment density operator is given by
\begin{equation}
\hat{\rho}_{SE}(t) = \hat{U}(t) (\hat{\rho}_S \otimes \ketbra{0_E}) \hat{U}^{\dagger}(t).
\end{equation}

Then, performing a projective measurement of the environment observable $\hat{O}_E$ with eigenbasis $\{ \ket{n_E} \}$, the $\dim_E$ possible results are given by Eq.~\eqref{eq:probability} with
\begin{equation}
\hat{M}_n (t) = \bra{n_E} \hat{U}(t) \ket{0_E}
\label{eq:krausoperator}
\end{equation}
acting only on $\mathcal{H}_S$. $\hat{\rho}_{S|n} (t)$ of Eq.~\eqref{eq:probability} is the normalized density operator on $\mathcal{H}_S$ after the measurement yields result $n$.

In the case of a mixed initial environment state $\hat{\rho}_E = \sum_{\tilde{n}} p_{\tilde{n}} \ketbra{\tilde{n}_E}$ diagonal in a different orthonormal basis $\{ \ket{\tilde{n}_E} \}$, the operator $\hat{M}_{n\tilde{n}}(t) = \sqrt{p_{\tilde{n}}} \bra{n_E} \hat{U}(t) \ket{\tilde{n}_E}$ describes one specific transition. If one remains agnostic about the initial state of the environment, the unnormalized numerator $\hat{M}_{n\tilde{n}}(t) \hat{\rho}_S \hat{M}_{n\tilde{n}}^{\dagger}(t)$ and the probability $p_{n\tilde{n}}(t)$ in Eq.~\eqref{eq:probability} must each be summed over $\tilde{n}$ before normalization, which generally maps a pure system state to a mixed state $\hat{\rho}_{S|n}(t)$. While any mixed state can be purified to a pure state $\ket{0_E}$ within a larger Hilbert space \cite{nielsen}, describing the measurement outcome $n$ by a single operator $\hat{M}_n(t)$ as in Eq.~\eqref{eq:krausoperator} requires that the purifying degrees of freedom also be included in the projective measurement (unless the $\hat{M}_{n \tilde{n}}(t)$ for all $\tilde{n}$ are proportional to each other). Keeping this restriction in mind, we will from now on assume that the environment is initially in the pure state $\ket{0_E}$.

If the environment is measured after the time evolution but the result is not registered (i.e., the only information the observer has is that a measurement took place), the density operator of the system is given by the weighted average
\begin{equation} \label{eq:unread}
\hat{\rho}_{S|} (t) \coloneqq \sum_{n} p_n (t) \hat{\rho}_{S|n} (t) = \sum_{n} \hat{M}_{n} (t) \hat{\rho}_S \hat{M}_{n}^{\dagger} (t) = \Tr_E \hat{\rho}_{SE}(t) \equiv \hat{\rho}_S(t),
\end{equation}
which is called an unread generalized measurement. It becomes apparent that the time evolution of the reduced density operator $\hat{\rho}_S$ can be interpreted as an unread generalized measurement after the joint unitary time evolution on $\mathcal{H}_{SE}$. This interpretation is very general, as formalized by the Stinespring dilation theorem \cite{Stinespring}.

\begin{theorem}[Stinespring dilation]
\label{thm:stinespring}
Let $\mathcal{E}$ be a completely positive trace-preserving (CPTP) map representing the dynamical evolution of a system's density operator $\hat{\rho}_S$ on a Hilbert space $\mathcal{H}_S$. Then, there exists an ancillary environment Hilbert space $\mathcal{H}_E$, an initial pure state $\ket{0_{E}} \in \mathcal{H}_E$, and a global unitary operator $\hat{U}$ acting on $\mathcal{H}_S \otimes \mathcal{H}_E$ such that the map can be expressed as
\begin{equation*}
    \mathcal{E}(\hat{\rho}_S) = \Tr_E \left( \hat{U} (\hat{\rho}_S \otimes \ketbra{0_E}) \hat{U}^{\dagger} \right) = \sum_{n} \hat{M}_{n} \hat{\rho}_S \hat{M}_{n}^{\dagger}
\end{equation*}
with $\hat{M}_n$ from Eq.~\eqref{eq:krausoperator}.
\end{theorem}
The theorem formalizes the intuitive idea that time evolution fundamentally is a unitary operation, and through inclusion of a large enough environment, the unitarity can be restored. 

In order for the unitary $\hat{U}$ in Theorem~\ref{thm:stinespring} to be given by the physical time evolution operator of the system-environment composite as in Eq.~\eqref{eq:krausoperator}, one must demand that the system and environment initially factorize as a product state $\hat{\rho}_S (0) \otimes \ketbra{0_E}$. Otherwise, the product input appearing in the theorem does not describe the actual initial state of the physical system-environment composite. A decomposition of a given CPTP map as in Eq.~\eqref{eq:unread} is referred to as a Kraus representation, and the $\hat{M}_n(t)$ as Kraus operators \cite{nielsen, Kraus}.

Post-selection is the process in which all experimental runs but those with a certain outcome are discarded. This conditional restriction influences a priori how the classical probabilities in the density operator change throughout the time of the experiment. Mathematically, the unread generalized measurement of Theorem~\ref{thm:stinespring} is replaced by Eq.~\eqref{eq:probability} with time-dependent $p_i (t)$, $\hat{\rho}_{S|i}(t)$ and $\hat{M}_i(t)$ if the environment is measured at time $t$ and only runs with result $i$ are kept. Note that the equation of motion for $\hat{\rho}_{S|i}(t)$ becomes nonlinear due to the division by the probability. This nonlinearity, however, need not be included when solving the equation for the time evolution of the post-selected state. Rather, one can normalize the state when calculating expectation values, with the understanding that the environment was measured at time $t$ and the runs with the wrong results were discarded. The effective time evolution operator acting on the system's states before normalization is then given by
\begin{equation} \label{eq:efftimeevo}
\hat{M}_i(t) \eqcolon \hat{U}^u_{\mathrm{eff}}(t).
\end{equation}
In particular, if and only if the unnormalized time evolution factorizes as 
\begin{equation} \label{eq:stationarityKraus}
\hat{M}_i (t) \hat{\rho}_S(0) \hat{M}_i^{\dagger}(t) = \Tr \left( \hat{M}_i^{\dagger}(t) \hat{M}_i(t)  \hat{\rho}_S (0) \right) \hat{\rho}_S (0),
\end{equation}
the initial state is stationary under post-selected dynamics.

From Eq.~\eqref{eq:krausoperator} and using $\hat{U}(0) = \hat{1}$ as well as the orthonormality of the basis set $\{ \ket{n_E} \}$, one concludes that only the branch $\hat{M}_0(t)$ fulfills the correct initial condition $\hat{M}_0(0) = \hat{1}$ necessary to be represented as a time evolution operator generated by a non-Hermitian Hamiltonian, i.e.,
\begin{equation} \label{eq:timeevooperator}
\hat{U}^u_{\mathrm{eff}}(t) = \hat{M}_0(t) = \mathcal{T} e^{-i \int_0^{t} \! dt' \, \hat{H}_{\mathrm{eff}}(t')}
\end{equation}
with the time-ordering symbol $\mathcal{T}$.

The time derivative of Eq.~\eqref{eq:timeevooperator} is $\partial_t \hat{U}^u_{\mathrm{eff}}(t) = -i \hat{H}_{\mathrm{eff}}(t) \hat{U}^u_{\mathrm{eff}}(t)$. Therefore, if and only if $\hat{M}_0(t)$ is invertible on the whole interval $[0, t]$, the effective Hamiltonian is given by
\begin{equation} \label{eq:exacteffhamiltonian}
\hat{H}_{\mathrm{eff}} (t) = i \left( \partial_t \hat{M}_0(t) \right) \hat{M}_0^{-1}(t).
\end{equation}
The sufficiency of the invertibility of $\hat{M}_0(t)$ follows by inserting the explicit construction of Eq.~\eqref{eq:exacteffhamiltonian} together with the initial condition into Eq.~\eqref{eq:timeevooperator}. The necessity arises because the desired form of $\hat{M}_0(t)$ in Eq.~\eqref{eq:timeevooperator} has the explicit inverse $ \hat{M}_0^{-1}(t) = \tilde{\mathcal{T}} e^{+i \int_0^{t} \! dt' \, \hat{H}_{\mathrm{eff}}(t')}$, where $\tilde{\mathcal{T}}$ is the anti-time-ordering symbol.

$\hat{M}_0(t)$ is always invertible for short enough times since it is a smooth function of time and approaches $\hat{1}$ as $t \to 0$. For a given system-environment unitary $\hat{U}(t)$, however, the resulting $\hat{M}_0(t)$ may become non-invertible at a finite time. Conversely, if one demands time evolution with a given $\hat{H}_{\mathrm{eff}}(t)$, finding a corresponding $\hat{U}(t)$ may only be possible up to a maximum time $\tau_{\mathrm{max}}$ because the probability interpretation of Eq.~\eqref{eq:probability} breaks down at later times. This second statement will be made more precise in the presentation of the ancilla approach in Sec.~\ref{subsec:ancillaapproach}; see also Ref.~\cite{PTpaper}.

Equation~\eqref{eq:exacteffhamiltonian} was already obtained in Ref.~\cite{InstantaneousSelfEnergy} for the single-particle case and used to define an ``instantaneous self-energy'' $\hat{\Sigma}_{\mathrm{inst}}(t)$. The latter is the correction to the projection $\ev{\hat{H}(t)}{0_E}$ of the full system-environment Hamiltonian $\hat{H}(t)$ generating $\hat{U}(t)$ of Theorem~\ref{thm:stinespring}. Therefore, we have
\begin{equation} \label{eq:instantaneousselfenergy}
\hat{H}_{\mathrm{eff}}(t) = \ev{\hat{H}(t)}{0_E} + \hat{\Sigma}_{\mathrm{inst}}(t).
\end{equation}
Equation~\eqref{eq:instantaneousselfenergy} is the exact, time-local dynamical generator of the post-selected time evolution before normalization. It contains all possible effects of gain, loss, and backaction from the environment. 

In the next section, we will draw a connection between the non-Hermitian Hamiltonians obtained from the Lindblad, Feshbach, and ancilla approaches and this exact form. In particular, we will show that the special case of a time-independent $\hat{H}_{\mathrm{eff}}(t) \equiv \hat{H}_{\mathrm{eff}}$ can either arise by approximating $\hat{\Sigma}_{\mathrm{inst}}(t)$ as constant in time when $\hat{H}(t) \equiv \hat{H}$ is constant (Feshbach and Lindblad approaches), or by delicately engineering $\hat{H}(t)$ in a manner that cancels all time dependence in Eq.~\eqref{eq:instantaneousselfenergy}, even when both terms are time-dependent (ancilla approach).

From now on, we denote the system density operator by $\hat{\rho}$, omitting the subscript $S$. Similarly, we omit the subscript specifying the measurement outcome of the generalized measurement as in Eq.~\eqref{eq:probability}, instead stating explicitly whenever the branch $\hat{M}_0(t)$ is post-selected for.

\section{Approaches to non-Hermitian dynamics} \label{sec:approaches}
In this section, effective non-Hermitian Hamiltonians are derived from Feshbach projection, the Lindblad master equation, and the ancilla approach. We mention which approximations are involved compared to the exact dynamics generated by the Hamiltonian $\hat{H}_{\mathrm{eff}}$ of Eq.~\eqref{eq:instantaneousselfenergy}, as well as the properties of the effective non-Hermitian Hamiltonians specific to each of the approaches. 

We note that recent literature has proposed schemes to access non-Hermitian dynamics without explicit post-selection protocols \cite{YanNoPostSelection, KujiNoPostSelection}. While these approaches provide experimentally relevant constructions for extracting non-Hermitian transition amplitudes, presenting post-selected dynamics within the Kraus-map framework remains useful as a theoretical unification. Importantly, in these post-selection-free protocols, the physical system state continues to evolve according to a standard trace-preserving Lindbladian. The non-Hermitian dynamics instead emerge at the level of specific expectation values: through an engineered non-Hermitian subspace of the Lindbladian operator space in Ref.~\cite{YanNoPostSelection}, or through statistical cancellation of unwanted jump contributions using stochastic quantum error mitigation in Ref.~\cite{KujiNoPostSelection}. These approaches thus avoid explicit post-selection, but do not generically eliminate exponential sampling costs. 

In a different approach \cite{ZhangNoPostSelection}, it was demonstrated that this complication can be entirely avoided under certain circumstances by simulating non-Hermitian dynamics using variational quantum circuits. However, the same work established that simulating the time evolution of highly entangled random states can require an exponentially large number of quantum gates even at short evolution times. As we will also show, the post-selection probability remains constant or fluctuates for linear combinations of eigenstates with real eigenvalues; it is thus expected that one can in principle delicately engineer initial states or Hamiltonians for which no exponential overhead occurs. Moreover, dissipative terms localized within a spatial region and proportional to the density bound the exponential decay rate of the post-selection probability in fermionic systems due to the Pauli exclusion principle. Hence, the decay rate does not necessarily grow with the system size. 

We therefore argue that the unifying picture presented here should not be understood as asserting that post-selection is universally the only or even the optimal route to simulating non-Hermitian dynamics. Rather, post-selection provides a general operational mechanism through which a non-unitary effective evolution emerges from an underlying quantum process, and the resulting success probability provides one natural measure of the associated simulation cost. Alternative protocols may avoid this particular overhead by shifting the computational burden to statistical estimation, circuit depth, or state complexity. The actual efficiency therefore depends on the structure of the initial state, the effective Hamiltonian, and the observable of interest.

\subsection{Feshbach projection} \label{subsec:feshbach}
The Feshbach projection technique is applicable whenever the full Hilbert space can be decomposed into a direct sum $\mathcal{H} = \hat{P} \mathcal{H} \oplus \hat{Q} \mathcal{H}$ with projectors $\hat{P}$ and $\hat{Q} = \hat{1} - \hat{P}$, and one is only interested in the reduced dynamics of states within the $\hat{P}$-subspace \cite{Feshbach2}. The choice of the projected subspace is entirely general, as the Feshbach technique provides an operator identity for arbitrary Hilbert spaces $\mathcal H$ and projectors $\hat P$.

In the present case, we focus on the system-environment decomposition $\mathcal{H}_1 = \mathcal{H}_{1,S} \oplus \mathcal{H}_{1, E}$ of the single-particle Hilbert space $\mathcal{H}_1$ which is relevant to the open-system context in which non-Hermitian Hamiltonians are typically engineered. Therein, a properly chosen orthonormal single-particle basis set $\{ \ket{i} \in  \mathcal{H}_1 \}$ can be segmented into states $\{ \ket{i^S} \}$ localized in the system and $\{ \ket{i^E} \}$ localized in the environment. As we are interested in many-body applications, we work in Fock space. The annihilation operators of the system and environment excitations are then given by $\{ \hat{c}_i^S \}$ and $\{ \hat{c}_i^E \}$, respectively. The full Fock space then possesses the tensor-product structure
\begin{equation}
\mathcal{F}(\mathcal{H}_{1, S} \oplus \mathcal{H}_{1, E}) \simeq \mathcal{F}(\mathcal{H}_{1, S}) \otimes \mathcal{F}(\mathcal{H}_{1, E}).
\end{equation}
In order to fulfill the correct (anti-)commutation relations between system and environment operators, the ladder operators on this tensor-product space need to be defined as
\begin{equation} \label{eq:ladderoperators}
\hat{b}_i^S  \coloneqq \hat{c}_i^S \otimes \hat{1}_E, \qquad \hat{b}_i^E \coloneqq (\pm1)^{\hat{N}_S} \otimes \hat{c}_i^E,
\end{equation}
where $\hat{N}_S$ counts the number of particles in the system and the upper (lower) sign describes bosons (fermions).

One then chooses a pure initial reference state $\ket{0_E}$ of the environment at $t=0$ and defines the projector
\begin{equation} \label{eq:projector}
\hat P=\hat 1_S\otimes\ketbra{0_E},
\end{equation}
such that the $\hat P$-subspace consists of tensor products between system states and the environment reference state. 

Note that the physical environment can generally be in a statistical mixture, e.g., a thermal state. In such a case, $\mathcal{H}_{1, E}$ is understood to include auxiliary degrees of freedom, so that the initial mixture is represented by the purified state $\ket{0_E}$ in the corresponding Fock space. Consequently, the operators $\{ \hat{c}_i^E \}$ need not represent physical excitations. As mentioned in the discussion after Eq.~\eqref{eq:krausoperator}, this extension is only warranted when the auxiliary degrees of freedom can be measured for post-selection purposes, unless the corresponding Kraus operators are proportional to each other.

Projection with $\hat{P}$ at time $t$ selects the component of the evolved state in which the environment has returned to its initial state. It can be understood as a projection onto the system's degrees of freedom since the environmental ones are fixed. If the total particle number is conserved, projection onto the initial environment state restores the system's initial particle number, provided the latter is sharply defined and the physical environment state is diagonal in particle number. Particles may nevertheless be exchanged during the evolution.

Conversely, post-selecting by measuring whether the system has retained its initial particle number generally does not ensure that the environment has returned to $\ket{0_E}$, as it may remain in an excited state with the same particle number. However, if the environment is initially in its vacuum state with respect to the physical excitations, the environmental subspace at that particle number is one-dimensional, so finding the initial particle number in the system at time $t$ guarantees that the environment has returned to $\ket{0_E}$. 

In the single-particle case, the direct-sum structure gives this post-selection a particularly simple interpretation: it selects the component of the evolved state in which the particle is found within the system at time $t$. This was experimentally realized by Naghiloo et al.\ using the two excited levels of a three-level superconducting transmon as an effective qubit (``system''), with the ground state acting as a loss channel (``environment'') \cite{Naghiloo}.

Denoting block-projected operators by $\hat{O}_{AB} \coloneqq \hat{A} \hat{O} \hat{B}$ for $\hat{A}, \hat{B} \in \{\hat{P}, \hat{Q}\}$, Feshbach showed that regardless of the choice of $\hat{P}$, the energy-dependent effective Hamiltonian for the system alone fulfills the relation
\begin{equation} \label{eq:feshbach}
\hat{H}_{\mathrm{eff}} (z) = \hat{H}_{PP} + \hat{\Sigma}(z), \qquad
\hat{\Sigma}(z) = \hat{H}_{PQ} \frac{1}{z - \hat{H}_{QQ}} \hat{H}_{QP}.
\end{equation}
$\hat{H}$ is the full, time-independent many-body Hamiltonian of both the system and the environment and $z$ denotes the variable in the complex energy plane. While $\hat{H}_{\mathrm{eff}}(z)$ and $\hat{\Sigma}(z)$ formally act on the entire Fock space, we identify the system-projected operators in the following with operators acting only on the system Fock space by omitting the block indicator $\otimes \ketbra{0_E}$.

In order to apply $\hat{H}_{\mathrm{eff}}(z)$, we calculate the Laplace transform of the retarded time evolution operator $\hat{G}^R(t) \coloneqq -i \Theta(t) e^{-i \hat{H} t}$ for complex energies $z$ in the upper half of the complex plane, which yields the resolvent 
\begin{equation} \label{eq:resolvent}
\hat{G}(z) = \int \! dt \, e^{izt} \hat{G}^R(t) = \frac{1}{z - \hat{H}}.
\end{equation}
The time independence of $\hat{H}$ is necessary in the Feshbach formalism to obtain the resolvent structure. Using the Feshbach projection, the block-projected resolvent consequently takes the form
\begin{equation} \label{eq:blockprojectedresolvent}
\hat{G}_{PP}(z) = \frac{1}{z - \hat{H}_{\mathrm{eff}}(z)}.
\end{equation}
When evaluated slightly above the real axis at $z = \omega + i0^+$, $\hat{G}(z)$ and $\hat{G}_{PP}(z)$ represent the full and projected retarded Green's functions $\hat{G}^R(\omega)$ and $\hat{G}_{PP}^R(\omega)$, respectively. The corresponding retarded self-energy is defined as $\hat{\Sigma}^R(\omega) \coloneqq \hat{\Sigma}(\omega + i 0^+)$. It is important to distinguish $\hat{\Sigma}(z)$ and $\hat{\Sigma}^R(\omega)$ from the instantaneous self-energy $\hat{\Sigma}_{\mathrm{inst}}(t)$ defined in Eq.~\eqref{eq:instantaneousselfenergy} as they are different mathematical objects. The connection between the two is given in Ref.~\cite{InstantaneousSelfEnergy} for $t>0$ as
\begin{equation}
\hat{\Sigma}_{\mathrm{inst}}(t) = \left[ \int_{-\infty}^{\infty} \! d\omega\, \hat{\Sigma}^R(\omega) \hat{G}_{PP}^R(\omega) e^{-i\omega t} \right] \left[ \int_{-\infty}^{\infty} \! d\omega\, \hat{G}_{PP}^R(\omega) e^{-i\omega t} \right]^{-1},
\label{eq:ConnectionSelfEnergies}
\end{equation}
making it clear that $\hat{\Sigma}_{\mathrm{inst}}(t)$ is not the inverse Fourier transform of $\hat{\Sigma}^R(\omega)$.

In the time domain, the projection of $\hat{G}^R(t)$ with Eq.~\eqref{eq:projector} results in 
\begin{equation} \label{eq:greensfct}
\hat{G}_{PP}^R(t) = - i \Theta(t) \ev{e^{-i \hat{H} t}}{0_E} = -i \Theta(t) \hat{M}_0(t),
\end{equation}
where we have identified the Kraus operator for the environment starting and ending up in its reference state at times $0$ and $t$, respectively (cf.\ Eq.~\eqref{eq:krausoperator}). 

Drawing from the formalism of Sec.~\ref{sec:generalizedmeasurements}, this means that the Green's function with the effective Hamiltonian includes just one unnormalized branch of the CPTP map for the system's reduced density matrix. In the usual application of the Feshbach formalism, no post-selection is performed, and $\hat{G}_{PP}(z)$ is used to extract properties of the system such as transition amplitudes or decay poles \cite{CohenTannoudji}. However, $\hat{G}_{PP}^R(t)$ does not by itself describe the unconditional time evolution of the reduced density matrix. Because the system can become entangled with the environment, it instead evolves through the sum over all Kraus branches, as specified in Theorem~\ref{thm:stinespring}.

In order to obtain genuine non-Hermitian time evolution, a necessary condition is to post-select for the branch $\hat{M}_0(t)$ of the CPTP map. By ``genuine'', we mean that quantum states are evolved deterministically by a time evolution operator and thus obey Eq.~\eqref{eq:schroedinger}. Pure states remain pure in the process. Whenever the post-selection and measurements of an observable in the system are performed (simultaneously), the system state is normalized.

Transforming Eq.~\eqref{eq:blockprojectedresolvent} back into the time domain results in the integro-differential equation
\begin{equation} \label{eq:timedomaingreens}
(i\partial_t-\hat{H}_{PP})\hat{G}^R_{PP}(t) - \int\! dt'\, \hat{\Sigma}^R(t-t')\hat{G}^R_{PP}(t') = \delta(t) \hat{1}_S,
\end{equation}
where the convolution of the retarded self-energy with the Green's function encodes the memory kernel of the environment. 

It is apparent that the time-nonlocal action of the retarded self-energy in the time domain encodes memory effects differently than the instantaneous self-energy of Eq.~\eqref{eq:instantaneousselfenergy}. In the latter, gain and loss are represented by anti-Hermitian terms, while $\hat{\Sigma}^R(\omega)$ need not be non-Hermitian to include these effects through the convolution. Indeed, for a discrete environment, the retarded self-energy is Hermitian apart from isolated real-axis singularities, as can be seen by inserting $z=\omega + i0^+$ in Eq.~\eqref{eq:feshbach}. In order to generate a non-Hermitian retarded self-energy inducing irreversible decay, the spectrum of the environment block $\hat{H}_{QQ}$ needs to be continuous, which is the case for environments with an infinite number of degrees of freedom (thermodynamic limit). Then, $\hat{\Sigma}^R(\omega)$ gains a finite anti-Hermitian part inside the continuum:
\begin{equation}\label{eq:selfenergy}
\hat{\Sigma}^{R}(\omega) = \mathcal{P} \int \! dE \, \frac{\hat{J}(E)}{\omega - E} - i \pi \hat{J}(\omega), 
\end{equation} 
where the system-projected spectral density of the environment is defined as
\begin{equation} \label{eq:spectraldensity}
\hat{J}(E) = \hat{H}_{PQ} \delta(E - \hat{H}_{QQ}) \hat{H}_{QP}.
\end{equation}
Since $\hat{J}(\omega)$ is a positive semi-definite operator, all $\omega$-dependent eigenvalues of $\hat{\Sigma}^R(\omega)$ have non-positive imaginary parts. Physically, this describes amplitude leaking from the $\hat{P}$ sector into environmental eigenstates with matching energy $\omega$. The coupling operators $\hat{H}_{PQ}$ and $\hat{H}_{QP}$ in Eq.~\eqref{eq:spectraldensity} mediate this leakage. For a finite environment the system amplitude can in principle revive due to the discrete spectrum; in the thermodynamic limit the level spacing vanishes, the recurrence time is pushed to infinity, and the projected dynamics becomes effectively irreversible. The principal value term in Eq.~\eqref{eq:selfenergy} includes virtual excursions of particles into environment states of mismatched energy and causes a shift of the system's energy levels.

If the self-energy is independent of $\omega$ or can be approximated as such, Eq.~\eqref{eq:timedomaingreens} turns into the time-local Schr\"odinger equation
\begin{equation} \label{eq:nonhermschroedinger}
i \partial_t \ket{\Psi^u (t)} = \hat{H}_{\mathrm{eff}} \ket{\Psi^u (t)} \quad \Rightarrow \quad \hat{M}_0 (t) = \hat{U}_{\mathrm{eff}}^u(t) = e^{-i \hat{H}_{\mathrm{eff}}t}
\end{equation}
with $\hat{H}_{\mathrm{eff}} = \hat{H}_{PP} + \hat{\Sigma}^R$. An $\omega$-independent self-energy means that the environment becomes memoryless because $\hat{\Sigma} (t-t') \propto \delta(t-t')$. This is known as the wide-band limit (WBL) since it is a consequence of the spectral density operator $\hat{J} (\omega) \equiv \hat{J}_0$ being constant over a large range in frequency space, making both terms in Eq.~\eqref{eq:selfenergy} approximately $\omega$-independent. In Ref.~\cite{InstantaneousSelfEnergy}, it was shown that in this special case, $\hat{\Sigma}_{\mathrm{inst}} = \hat{\Sigma}^R$ coincide as expected. As we will see in Sec.~\ref{subsec:lindblad}, the memoryless environment is similar to the Markov approximation, which assumes a separation of time scales between the memory of the bath and the time scale on which the system evolves.

Strictly speaking, an $\omega$-independent spectral density over the entire real axis would imply that the principal value part of Eq.~\eqref{eq:selfenergy} vanishes, meaning that the self-energy is purely anti-Hermitian and no level shifts occur. Often, the WBL is therefore understood to approximate $\hat{\Sigma}(\omega)$ by its value at a resonance energy $\epsilon_0$ of the system, also allowing for a Hermitian correction to $\hat{H}_{PP}$ \cite{Stefanucci}.

Although it neglects a variety of memory-induced transient and backaction effects included in the instantaneous self-energy of Eq.~\eqref{eq:instantaneousselfenergy} (cf. Ref.~\cite{InstantaneousSelfEnergy}), the WBL approximation is useful since a static Hamiltonian allows the dynamics to be analyzed in terms of eigenstates and eigenvalues. For this reason, the literature has in large parts been focused on time-independent non-Hermitian Hamiltonians. As mentioned before, the time independence of $\hat{H}_{\mathrm{eff}}$ can be enforced without an approximation, which will be shown in Sec.~\ref{subsec:ancillaapproach}. However, the Feshbach construction does not rely on a fine-tuned construction of the system-environment dynamics and is thus an experimentally accessible way to obtain non-Hermitian Hamiltonians.

With the WBL in place, virtual excursions of particles into the environment must be effectively instantaneous; a particle cannot leave the system, linger in the environment for a finite amount of time, and then return. An alternative picture to illustrate this consequence is the quantum trajectory picture. The events happening during a trajectory are called quantum jumps, which are instantaneous transfers of particles or holes from the system to the environment. More generally, quantum jumps can also describe environment-mediated changes of the system that do not involve particle transfer. If one were able to continuously monitor these events, one would obtain a history of the composite system in terms of such jumps. In fact, Dehmelt and coworkers \cite{Dehmelt} famously demonstrated experimentally that instantaneous transitions between energy levels can be observed using a trapped Ba$^+$ ion. In the present context, all quantum jumps are forbidden when post-selecting for the reference sector of the environment. This is because the probability of a subsequent jump restoring the environmental reference state tends to zero when the environment is memoryless and has infinitely many degrees of freedom. 

The post-selection success probability for an initial system state $\ket{\Psi(0)}$ following the conditional dynamics of Eq.~\eqref{eq:nonhermschroedinger} is given by 
\begin{equation} \label{eq:postselectionprobability}
p_0 (t) = \ev*{\hat{M}_0^{\dagger}(t) \hat{M}_0(t)}{\Psi (0)} = \ev{e^{i\hat{H}_{\mathrm{eff}}^{\dagger}t} e^{-i\hat{H}_{\mathrm{eff}}t}}{\Psi (0)},
\end{equation}
cf.\ Eq.~\eqref{eq:probability}.

If the initial state is a normalized right eigenstate $\ket{\Psi (0)} = \ket*{\Psi^R_i}$ of $\hat{H}_{\mathrm{eff}}$ with eigenvalue $E_i + i \Gamma_i$ and $\Gamma_i < 0$, then the probability decays exponentially, $p_0(t) = e^{2 \Gamma_i t}$. If the eigenvalue is purely real, $p_0(t) \equiv 1$. This means that $\ket*{\Psi^R_i} \otimes \ket{0_E}$ is an (approximate) eigenstate of the full Hamiltonian $\hat{H}$ and amplitude is never funneled into other environment sectors $\ket{n_E}$. In the context of quantum optics, such states are an example of so-called ``dark states'' with respect to the monitored decay channels since no photons are emitted into or absorbed from the environment \cite{DarkStates}. Regardless of the eigenvalue, all eigenstates of $\hat{H}_{\mathrm{eff}}$ are stationary under the time evolution due to the normalization imposed by the generalized measurement, see the denominator of Eq.~\eqref{eq:probability}. 

In contrast to that, for a coherent superposition
\begin{equation} \label{eq:superpositionstate}
\ket{\Psi(0)} = \frac{\ket*{\Psi_1^R}+\ket*{\Psi_2^R}}{\sqrt{2+2\operatorname{Re}\braket*{\Psi_2^R}{\Psi_1^R}}}
\end{equation}
of two normalized right eigenstates with eigenvalues $E_{1/2} + i \Gamma_{1/2}$, the post-selection success probability and time-evolved state are given by
\begin{align}
p_0(t) &= \frac{e^{2\Gamma_1 t}+e^{2\Gamma_2 t}+2e^{(\Gamma_1+\Gamma_2)t}\operatorname{Re}\!\left[e^{i(E_2-E_1)t}\braket*{\Psi_2^R}{\Psi_1^R}\right]}{2+2\operatorname{Re}\braket*{\Psi_2^R}{\Psi_1^R}}, \label{eq:probsuperp}\\
\ket{\Psi(t)} &= \frac{e^{(\Gamma_1-iE_1)t}\ket*{\Psi_1^R}+e^{(\Gamma_2-iE_2)t}\ket*{\Psi_2^R}}{\sqrt{\left(2+2\operatorname{Re}\braket*{\Psi_2^R}{\Psi_1^R}\right)p_0(t)}}, \label{eq:propsuperpstate}
\end{align}
respectively.
The post-selection probabilities of the individual states decay with rates $2 |\Gamma_i|$, which results for $t \to \infty$ in a projection onto the mode with the slower decay. In addition, there is also a sinusoidal fluctuation of $p_0(t)$ due to the interference of the two non-orthogonal eigenstates that decays with the mean of the two rates. As a general interpretation, the knowledge of the state surviving post-selection up to time $t$ updates the quantum mechanical probabilities in favor of the eigenstate which has a smaller loss rate, in addition to a coherent phase relationship causing transient fluctuations. 

To clearly distinguish the physical behaviors governing $p_0(t)$, we refer to the overall asymptotic growth or decay of the state, driven by the imaginary parts $\Gamma_i$ of the eigenvalues, as exponential gain or exponential loss. For intermediate times, the superposition of the exponential and sinusoidal terms drives fluctuations of $p_0(t)$. We refer to periods where this interference temporarily causes the total probability to increase ($\partial_t p_0(t) > 0$) as transient gain, and periods where it decreases ($\partial_t p_0(t) < 0$) as transient loss.

Examining Eq.~\eqref{eq:probsuperp}, one might suspect that the transient behavior could cause $p_0(t)$ to exceed unity. However, due to the negative semi-definite nature of $-i \hat{H}^a_{\mathrm{eff}}$ (cf.\ Eq.~\eqref{eq:selfenergy}), one finds that
\begin{equation} \label{eq:semidefinite}
\partial_t p_0(t) = -i p_0(t) \ev{\hat{H}_{\mathrm{eff}} - \hat{H}_{\mathrm{eff}}^{\dagger}}{\Psi(t)} = -2i p_0(t) \ev{\hat{H}^a_{\mathrm{eff}}}{\Psi(t)}\leq 0,
\end{equation} 
and the probability decreases monotonically. In the established terminology, neither transient nor exponential gain is possible.

As Eq.~\eqref{eq:propsuperpstate} shows, a coherent superposition of two right eigenstates is generally nonstationary. This is also true for Hermitian dynamics, where superpositions of eigenstates with different energies acquire time-dependent relative phases. In contrast, even statistical mixtures of right eigenstates can be nonstationary under non-Hermitian dynamics. Using the eigenstates from the previous example, one obtains
\begin{equation} \label{eq:mixturesevo}
\begin{aligned}
\hat{\rho}(0) &= p_1 \ketbra*{\Psi^R_1} + p_2 \ketbra*{\Psi^R_2},\\
\hat{\rho}(t) &= \frac{p_1 e^{2\Gamma_1 t}}{p_1 e^{2\Gamma_1 t} + p_2 e^{2\Gamma_2 t}} \ketbra*{\Psi^R_1} + \frac{p_2 e^{2\Gamma_2 t}}{p_1 e^{2\Gamma_1 t} + p_2 e^{2\Gamma_2 t}} \ketbra*{\Psi^R_2},
\end{aligned}
\end{equation}
which maps in the $t \to \infty$ limit to the longer-lived mode. The evolution only affects the diagonal populations, i.e., no coherences build up if the initial density operator is diagonal in the non-orthogonal right eigenbasis. Only statistical mixtures of eigenstates with identical $\Gamma$ are stationary ensembles (this will be derived on more general grounds in Sec.~\ref{sec:equilibrium}).

Because all eigenvalues of $\hat{H}_{\mathrm{eff}}$ generated through Feshbach projection have non-positive imaginary parts, \PT-symmetric non-Hermitian Hamiltonians cannot be obtained in this way. In the so-called \PT-broken phase, defined by the presence of complex eigenvalues despite the \PT{} symmetry, eigenvalues always appear in complex conjugate pairs \cite{PTpaper}. This would necessitate eigenvalues with positive imaginary parts. In the \PT-unbroken phase, all eigenvalues of $\hat{H}_{\mathrm{eff}}$ are strictly real. For this to be true, the positive semi-definite operator $\hat{J}$ in Eq.~\eqref{eq:selfenergy} has to be identically $\hat{0}$, which would make the Hamiltonian Hermitian. 

An additional restriction arises from the negative semi-definite nature of $-i\hat{H}_{\mathrm{eff}}^a$. If a right eigenstate $\ket*{\Psi^R_i}$ possesses a purely real eigenvalue $E_i$, one obtains $\operatorname{Im}(E_i) = \ev*{-i\hat{H}_{\mathrm{eff}}^a}{\Psi^R_i} = 0$, which requires $\hat{H}_{\mathrm{eff}}^a \ket*{\Psi^R_i} = 0$. Consequently, $\ket*{\Psi^R_i}$ must also be an eigenstate of the Hermitian part $\hat{H}^h_{\mathrm{eff}} = \frac{1}{2} (\hat{H}_{\mathrm{eff}} + \hat{H}_{\mathrm{eff}}^\dagger)$ with eigenvalue $E_i$. For two such eigenstates, we derive the relation
\begin{equation} \label{eq:orthogonalrealevs}
0 = \bra*{\Psi^R_i} (\hat{H}_{\mathrm{eff}}^h - \hat{H}_{\mathrm{eff}}^h) \ket*{\Psi^R_j} = (E_i - E_j) \braket*{\Psi^R_i}{\Psi^R_j},
\end{equation}
where one $\hat{H}_{\mathrm{eff}}^h$ was applied to the ket and the other one to the bra. Therefore, the construction cannot produce non-orthogonal eigenstates with real eigenvalues except within a degenerate subspace.

These two limitations show how $-i \hat{H}_{\mathrm{eff}}^a$ being negative semi-definite strongly constrains the class of possible non-Hermitian Hamiltonians obtained via Feshbach projection. The constraint is due to the imposition of the WBL, which is necessary for time-local dynamics in the Feshbach formalism, in contrast to Eq.~\eqref{eq:instantaneousselfenergy}.   

One way to circumvent these restrictions is to use a passive embedding \cite{ornigotti, Joglekar}. Suppose that one wants to obtain a non-Hermitian target Hamiltonian $\hat{H}_{\mathrm{target}}$ that is not achievable through Feshbach projection due to the reasons discussed above. What one does instead is to construct a different Hamiltonian 
\begin{equation} \label{eq:passivegain}
\hat{H}_{\mathrm{eff}} = \hat{H}_{\mathrm{target}} - i \kappa \hat{1},
\end{equation}  
where the constant $\kappa > 0$ has to be chosen larger than the largest eigenvalue of $-i \hat{H}_{\mathrm{target}}^a$, meaning that the spectrum of the anti-Hermitian part needs to be bounded along the positive imaginary axis. We stress that it is not enough to choose $\kappa$ larger than the largest imaginary eigenvalue part of the full target Hamiltonian $\hat{H}_{\mathrm{target}}$. 

With the passive embedding, $-i \hat{H}_{\mathrm{eff}}^a$ is negative semi-definite and $\hat{H}_{\mathrm{eff}}$ can be realized. All post-selection success probabilities of $\hat{H}_{\mathrm{target}}$ are then measured relative to the global loss term $e^{-2 \kappa t}$. This notion of an exponential increase in $p_0(t)$ relative to the global loss term is called passive (exponential) gain, and we distinguish it from active gain, which is exponential gain in absolute terms and cannot arise from the Feshbach construction.

Because the eigenstates are not affected by the constant offset in Eq.~\eqref{eq:passivegain} and Eq.~\eqref{eq:nonlinschroedinger} is invariant under the shift $-i \kappa \hat{1}$, it is possible to study the properties of $\hat{H}_{\mathrm{target}}$ using $\hat{H}_{\mathrm{eff}}$. All mentioned restrictions on the nature of $\hat{H}_{\mathrm{target}}$ can thereby be lifted, as long as $-i\hat{H}_{\mathrm{target}}^a$ is bounded from above. Due to the asymptotic exponential loss, however, the number of successful experimental runs will exponentially approach zero for long times. 

An insufficient number of successful post-selections is not just a theoretical problem. In the aforementioned experimental realization of a non-Hermitian system by Naghiloo et al.~\cite{Naghiloo}, significant parameter regimes remained inaccessible to the quantum state tomography performed, as not enough experimental runs survived the post-selection.

In summary, only under enforced post-selection with respect to the initial purified environment state, together with a dense environmental spectrum and a memoryless environment in the WBL, does one obtain genuine non-Hermitian Hamiltonian dynamics from a Feshbach projection. The effective Hamiltonian is thereby a static approximation of the exact dynamics. No active gain and no \PT-symmetric Hamiltonians can be constructed this way, except through a passive embedding.

\subsection{Lindblad master equation} \label{subsec:lindblad}
The Lindblad master equation can be derived most naturally from Kraus maps (see also Ref.~\cite{harochebook}). Assuming a Markovian environment, the time scale $T$ on which the system density operator changes, $T \| \partial_t \hat{\rho} (t) \| \sim \| \hat{\rho} (t) \|$, is much larger than the time scale $\tau_c$ on which the memory kernel of the bath decays. Connecting this to Sec.~\ref{subsec:feshbach}, the effective spectral density of the bath (Eq.~\eqref{eq:spectraldensity}) changes on the energy scale $1/\tau_c$. The existence of a finite correlation time $\tau_c$ supporting an irreversible Markovian description for arbitrarily long times relies on the thermodynamic limit, in which recurrences are pushed to infinity. For a finite environment, recurrences eventually restore long-time correlations. We therefore find that the Markov approximation includes very similar assumptions to the ones used to derive non-Hermitian self-energies in the WBL in Sec.~\ref{subsec:feshbach}.

As a consequence of Markovianity, there is always an intermediate time scale $\tau$ such that 
\begin{equation} \label{eq:separationofscales}
\tau_c \ll \tau \ll T.
\end{equation}
$\tau$ is the coarse-grained infinitesimal time step in the derivation of the differential equation for $\hat{\rho}(t)$. 

An additional assumption is that the density operator of $\mathcal{H}_E$ is effectively unaffected by interactions with the system, which is known in the present context as the Born approximation \cite{BreuerPetruccione}. This assumption is intuitive if the environment is much larger than the system or the interaction between system and environment is weak. 

Initially, the system-environment composite is in the product state 
\begin{equation} \label{eq:globalstate}
\hat{\rho}_{SE}^{\mathrm{glob}} (0) = \hat{\rho} (0) \otimes \ketbra{0_E}^{\mathrm{glob}}.
\end{equation}
The superscripts indicate that the density operators characterize the global environment. Whenever they are omitted, it signifies that only the local degrees of freedom of the environment relevant to the interaction with the system are regarded. 

After one time step $\tau$, the joint density operator evolves into an entangled state which can be represented by the expansion $\hat{\rho}_{SE}^{\mathrm{glob}}(\tau) = \sum_{m,n} \hat{S}_{mn}(\tau) \otimes \ketbra*{m_E}{n_E}^{\mathrm{glob}}$. 

The irreversible information dissipation into the Markovian environment implies that the correlations in $\hat{\rho}_{SE}(\tau)$ between the system and the local environment subside within the interval $I \coloneqq [\tau, \tau + \tau_c]$. Afterward, the state factorizes again as $\hat{\rho}_{SE}(\tau + \tau_c) \approx \hat{\rho}(\tau + \tau_c) \otimes \hat{\rho}_{E}(\tau + \tau_c)$. Due to the Born approximation, one can approximate $\hat{\rho}_{E}(\tau + \tau_c) \approx \ketbra{0_E}$, i.e., the environment reverts back to its initial state. We emphasize again that these approximations concern only the local environmental degrees of freedom that continue to interact with the system, not the global environment state. The latter may retain emitted excitations, remain correlated with the system and become orthogonal to its initial state $\ket{0_E}$. In particular, it does not mean that $\hat{\rho} (\tau + \tau_c) = \hat{S}_{00} (\tau + \tau_c)$ in the expansion of the global environment above. Instead, $\hat{\rho} (\tau + \tau_c) = \Tr_E \hat{\rho}_{SE}^{\mathrm{glob}}(\tau + \tau_c) = \sum_n \hat{S}_{nn} (\tau + \tau_c)$; but because the correlated, interacting dynamics of the $\hat{S}_{nn}$ are highly nontrivial, another application of the Markov approximation is needed.

Because of the Markov condition $\tau \gg \tau_c$, most of the time evolution within the subsequent interval $[\tau, 2\tau]$ happens in the subinterval $[\tau + \tau_c, 2\tau]$ as long as the interaction strength is much smaller than $\frac{1}{\tau_c}$. Therefore, ignoring the interval $I$ completely and writing 
\begin{equation} \label{eq:bornmarkovfull}
\hat{\rho}_{SE}(\tau) \approx  \hat{\rho}_{SE}(\tau + \tau_c) \approx \hat{\rho}(\tau + \tau_c) \otimes \ketbra{0_E} \approx \hat{\rho} (\tau) \otimes \ketbra{0_E}
\end{equation}
as the initial state of the second time step is valid. This is the crucial implication of the Born-Markov approximation, as it means that the necessary factorization as discussed after Theorem~\ref{thm:stinespring} is fulfilled at each time step and one can find infinitesimal Kraus operators $\{ \hat{M}_i(\tau) \}$ depending only on $\tau$ but not on $t$. It is what allows for the construction of a differential equation for $\hat{\rho}(t)$ alone.

While the approximations used here are similar to those in Sec.~\ref{subsec:feshbach}, the physical situation differs substantially in that the environment undergoes a repeated unread generalized measurement after each step $\tau$. This yields a temporal picture of full quantum trajectories, whereas in the Feshbach picture only the state of the environment at the time of measurement is relevant. 

A physically intuitive realization of an observed environment would be a quantum optical experiment in which detectors with a time resolution of $\tau$ constantly monitor whether or not a photon has escaped from a lossy cavity. A readout of the data unveils the exact quantum trajectory, while discarding the measurement record (or not using detectors in the first place) necessitates the ensemble description of the Lindblad equation. Viewing the time evolution of an ensemble as a weighted average over individual quantum trajectories is the foundational concept applied to the study of the Lindblad equation using Monte Carlo methods \cite{TrajectoryMonteCarlo}.

One can think of a time-dependent environmental Hilbert space, enlarging with each time step $\tau$. After the first step, it contains all one-step trajectories $\{ \ket{n_E}_{\tau}\}$, after the second one all two-step trajectories $\{ \ket{n_E}_{\tau} \otimes \ket{m_E}_{2\tau} \}$, and so on. The time evolution operator on the full ``temporal'' Hilbert space 
\begin{equation} \label{eq:temporalhilbertspace}
\mathcal{H}(t) = \mathcal{H}_S \otimes \bigotimes_{j = 1}^{t/\tau} \left( \mathcal{H}_E \right)_{j \tau}
\end{equation}
is given by the sequential interactions
\begin{equation}
\hat{U}(t) = \hat{U}_{SE_{t/\tau}}(\tau) \cdots \hat{U}_{SE_2}(\tau)\hat{U}_{SE_1}(\tau),
\end{equation}
where $\hat{U}_{SE_j}(\tau)$ acts only on the system and on $\left( \mathcal{H}_E \right)_{j \tau}$, but its action is the same for all $j$. This insight clarifies why the environment can be taken to start in the reference state $\ket{0_E}$ at the beginning of each time step $\tau$, even though the global environment may have evolved into a state orthogonal to its initial state: In contrast to the Feshbach formalism, the ``bookkeeping'' of possible quantum jumps is carried out by recording the outcomes at each time step rather than by determining the overlap between the initial and final environmental states. Consequently, the number of Kraus operators describing distinct quantum trajectories increases with each time step, with every operator corresponding to a specific sequence of jump outcomes.

Therefore, the ``no-jump'' Kraus operator means that no jump event occurred during the entire time evolution of the system, while in Sec.~\ref{subsec:feshbach} and before the imposition of the WBL, trajectories where, e.g., a particle left the system and an identical particle jumped back into it were allowed. With the WBL in place, however, both pictures are reconciled and yield the same results for the no-jump trajectory, cf.\ the discussion of quantum jumps following Eq.~\eqref{eq:nonhermschroedinger}.

We can now formulate the reduced dynamics over an infinitesimal time step $\tau$ in terms of Kraus operators. To leading order
in $\tau$, they can be written as
\begin{equation} \label{eq:krausansatz}
\hat{M}_0 = \hat{1} - i \hat{H}_{\mathrm{eff}} \tau \quad \text{and} \quad \hat{M}_{n \geq 1} = \sqrt{\tau} \hat{L}_{n}, 
\end{equation}   
with generic operators $\hat{H}_{\mathrm{eff}}$ and $\{ \hat{L}_n \}$. The latter are called ``jump operators'' because the corresponding process is a quantum jump from $\ket{0_E}$ to $\ket{n_E}$ happening during the time step $\tau$, i.e., quasi-instantaneously (cf.\ Eq.~\eqref{eq:krausoperator}).

From Eq.~\eqref{eq:krausoperator}, one would expect the first order of $\hat{M}_n$ to be linear in $\tau$. This would result in the first nonvanishing order of the transition probability being quadratic, i.e., a quantum-Zeno-type freezing of transitions. From Fermi's golden rule, however, it is known that a linear contribution arises due to the integration over a dense continuum of states, hence the square root leading order in $\tau$ for $n \geq 1$.

The completeness condition of Eq.~\eqref{eq:completeness} implies
\begin{equation} \label{eq:lindbladhamiltonian}
\hat{H}^a_{\mathrm{eff}} = -\frac{i}{2} \sum_{n \geq 1} \hat{L}_n^{\dagger} \hat{L}_n.
\end{equation}
The Hermitian part is not restricted further and contains the original system Hamiltonian as well as a so-called Lamb shift term, which in the language of Sec.~\ref{subsec:feshbach} is the Hermitian part of the self-energy. Equation~\eqref{eq:lindbladhamiltonian} shows that $-i \hat{H}^a_{\mathrm{eff}}$ is negative semi-definite, as in the Feshbach case (see Eq.~\eqref{eq:selfenergy}). We note that the isolated system Hamiltonian and hence also $\hat{H}_{\mathrm{eff}}$ can be time-dependent on a time scale much larger than $\tau$, possibly accompanied by time-dependent jump operators. This is in contrast to the Feshbach case, where the time independence was needed for the resolvent structure, cf.\ Eq.~\eqref{eq:resolvent}.

Finally, the Lindblad equation follows as the infinitesimal unread generalized measurement with the Kraus maps from Eq.~\eqref{eq:krausansatz},
\begin{equation} \label{eq:lindblad}
\partial_t \hat{\rho}(t) = -i (\hat{H}_{\mathrm{eff}}(t) \hat{\rho} (t) - \hat{\rho} (t) \hat{H}_{\mathrm{eff}}^{\dagger}(t)) + \sum_{n \geq 1} \hat{L}_n \hat{\rho} (t) \hat{L}_n^{\dagger}.
\end{equation}
Without the second term on the right-hand side, this expression is the generalized von Neumann equation 
\begin{equation} \label{eq:generalizedvonNeumann}
i \partial_t \hat{\rho}^u(t) = \hat{H}_{\mathrm{eff}}(t) \hat{\rho}^u (t) - \hat{\rho}^u (t) \hat{H}_{\mathrm{eff}}^{\dagger}(t),
\end{equation}
equivalent to the unnormalized (note the superscript on $\hat{\rho}^u(t)$) Schr\"odinger equation given in Eq.~\eqref{eq:schroedinger}.

The no-jump trajectory Kraus map is obtained by omitting the jump operator sum in Eq.~\eqref{eq:lindblad}. Integration up to time $t$ then leads to the finite-time result
\begin{equation} \label{eq:nojump}
\hat{M}_{0} (t) = \mathcal{T} e^{-i \int_0^t \! dt' \, \hat{H}_{\mathrm{eff}} (t')},
\end{equation}
which is in exact correspondence to Eq.~\eqref{eq:nonhermschroedinger} if $\hat{H}_{\mathrm{eff}}$ is time-independent.

In summary, the same class of effective Hamiltonians obtained through Feshbach projection with post-selection can also be derived in the Lindblad case, owing to the shared assumption of a Markovian environment in the thermodynamic limit. The essential difference is that the general Lindblad equation describes all possible trajectories. Nevertheless, only the no-jump trajectory is relevant for non-Hermitian dynamics in both cases.

\subsection{Ancilla approach} \label{subsec:ancillaapproach}
The third approach yielding non-Hermitian dynamics that is found in the literature is the ancilla approach \cite{AncillaOriginal, AncillaIntroduction, PTpaper, WuPTSymmetry}. In contrast to the other two formalisms, it does not assume the environment to be Markovian or have a dense spectrum. Instead, the environment Hilbert space is modeled by a simple spin-$1/2$ in which we label the $\hat{\sigma}_z$ eigenstates by $\ket{0_E}$ and $\ket{1_E}$ (spin-up and -down, respectively) to keep the notation consistent with previous sections. The possibility of amplitude revival in the system is crucial for achieving active gain. Furthermore, the full Hamiltonian $\hat{H}(t)$ can also be explicitly time-dependent. The exact dynamics of the composite system are then engineered to obtain a desired, general non-Hermitian Hamiltonian $\hat{H}_{\mathrm{eff}}(t)$ equivalent to Eq.~\eqref{eq:instantaneousselfenergy}.

We first note that one can reobtain the environments from the previous subsections by modifying the ancilla: By adding a register of multiple spin-$1/2$, the environment can be enlarged to model a bath of higher complexity, allowing also for a dense spectrum and irreversible loss. Further, the Markov approximation can be implemented at the level of a repeated-interaction or collision model: during each time step $\tau$, the system interacts with one ancillary bath unit, which is subsequently discarded and replaced by a fresh ancilla prepared in the reference state $\ket{0_E}$. This construction suppresses the memory of the environment and yields Markovian dynamics in the continuous-time limit \cite{ciccarello}. It was used and physically motivated through the Born-Markov approximation in our derivation of the Lindblad equation, see the discussion after Eq.~\eqref{eq:separationofscales}.

Keeping these possible generalizations in mind, we now turn to the study of a single-spin environment. At $t=0$, the system-ancilla composite is described by the state
\begin{equation} \label{eq:initialstateancilla}
\ket{\Psi_{SE}(0)} = \mathcal{N} \big(\ket{\Psi^u_{S} (0)} \otimes \ket{0_E} + \ket{\Psi^u_{S1} (0)} \otimes \ket{1_E}\big),
\end{equation}
with the normalization constant $\mathcal{N} = \big[ \braket{\Psi_S^u(0)} + \braket{\Psi_{S1}^u(0)} \big]^{-1/2}$ and unnormalized system kets $\ket{\Psi^u_{S}(0)}$ and $\ket{\Psi^u_{S1}(0)}$. The subscript of $\ket{\Psi^u_{S1}(0)}$ indicates that the system state belongs to the $\ket{1_E}$ sector of the environment, while the reference sector $\ket{0_E}$ is assigned no further label. One major difference compared to the other formalisms lies in the presence of initial probability in the non-post-selected sector. 

Then, a Hamiltonian $\hat{H}(t)$ acting on $\mathcal{H}_{SE}$ is constructed such that $\ket{\Psi^u_S(t)}$ evolves according to Eq.~\eqref{eq:schroedinger} with a given non-Hermitian Hamiltonian $\hat{H}_{\mathrm{eff}}(t)$ acting on $\mathcal{H}_S$. This is done by imposing the closure condition
\begin{equation} \label{eq:ancillaansatz}
\ket{\Psi_{S1}^u(t)} = \hat{g}(t) \ket{\Psi_S^u(t)}
\end{equation}
with an operator $\hat{g}(t)$ to be determined.

The result can be put in terms of projectors $\hat{P}$ and $\hat{Q}$ as in Sec.~\ref{subsec:feshbach}. By identifying the total Hilbert space as $\mathcal{H}_S \otimes \operatorname{span}\{\ket{0_E}, \ket{1_E}\} \cong \mathcal{H}_S \oplus \mathcal{H}_S$, we can represent states as two-component vectors and operators as $2 \times 2$ block matrices. The projectors $\hat{P}$ and $\hat{Q}$ select the upper and lower blocks, respectively. The closure condition then enforces that the Schr\"odinger equation for the $\ket{0_E}$ sector component takes the time-local form
\begin{equation} \label{eq:exactclosure}
i \partial_t \ket{\Psi_S^u(t)} = \left( \hat{H}_{PP}(t) + \hat{H}_{PQ}(t) \hat{g}(t) \right) \ket{\Psi_S^u(t)},
\end{equation} 
so that $\hat{H}_{\mathrm{eff}}(t) = \hat{H}_{PP}(t) + \hat{H}_{PQ}(t) \hat{g}(t)$. In contrast to the WBL self-energy discussed following Eq.~\eqref{eq:timedomaingreens}, time locality in Eq.~\eqref{eq:exactclosure} is not an approximation for a given system-environment Hamiltonian. Instead, it arises from the exact engineering of a Hamiltonian such that the solution is of the form of Eq.~\eqref{eq:ancillaansatz}, independent of the initial normalized system state $\ket{\Psi_S(0)}$. Then, the conditional dynamics of the system are generated by Eq.~\eqref{eq:instantaneousselfenergy} with the desired non-Hermitian Hamiltonian.

If the environment is larger and contains states $\{ \ket{n_E}, 1\leq n\leq N \}$, the full quantum state can initially possess weight in all sectors. Therefore, the closure condition extends to the existence of a set of $N$ operators $\{ \hat{g}_n(t): 1\leq n\leq N \}$ such that $\ket{\Psi_{Sn}^u(t)} = \hat{g}_n(t) \ket{\Psi_S^u(t)}$ for all $n$ and Eq.~\eqref{eq:exactclosure} is modified accordingly with a sum over all environmental block components. Solving the resulting set of coupled differential equations for the operators $\hat{H}(t)$ and $\hat{g}_n(t)$ quickly becomes analytically unfeasible. This is why the ``single-spin'' environment is routinely considered in the ancilla approach \cite{AncillaIntroduction, WuPTSymmetry}.

Equation~\eqref{eq:ancillaansatz} and the given effective Hamiltonian provide sufficient constraints to explicitly construct a family of solutions for $\hat{H}(t)$ and $\hat{g}(t)$. The result proposed in Refs.~\cite{WuPTSymmetry, PTpaper} is the coupled set of differential equations
\begin{align} \label{eq:ancillaresults}
    i\partial_t \mathbf{\hat{v}}(t) &= \mathbf{\hat{H}}(t) \mathbf{\hat{v}}(t) - \mathbf{\hat{v}}(t) \hat{H}_{\mathrm{eff}}(t), \\
    i\partial_t \hat{F}(t) &= \hat{H}_{\mathrm{eff}}^\dagger(t) \hat{F}(t) - \hat{F}(t) \hat{H}_{\mathrm{eff}}(t), \label{eq:ancillaresults2}
\end{align}
where the operators are defined as
\begin{equation} \label{eq:defops}
\begin{aligned}
\mathbf{\hat{H}}(t) &\coloneqq \begin{pmatrix} \ev{\hat{H}(t)}{0_E} & \mel{0_E}{\hat{H}(t)}{1_E} \\ \mel{1_E}{\hat{H}(t)}{0_E} & \ev{\hat{H}(t)}{1_E} \end{pmatrix}, \\
\mathbf{\hat{v}}(t) &\coloneqq \begin{pmatrix} \hat{1} \\ \hat{g}(t) \end{pmatrix}, \qquad \hat{F}(t) \coloneqq \mathbf{\hat{v}}^\dagger(t) \mathbf{\hat{v}}(t).
\end{aligned}
\end{equation}
This procedure possesses solutions valid for all $t$ if $\hat{H}_{\mathrm{eff}}$ is a diagonalizable pseudo-Hermitian Hamiltonian with only real eigenvalues. Pseudo-Hermiticity refers to the existence of a Hermitian invertible operator $\hat{\eta}$ such that 
\begin{equation} \label{eq:pseudohermiticity} 
\hat{H}_{\mathrm{eff}}^\dagger = \hat{\eta} \hat{H}_{\mathrm{eff}} \hat{\eta}^{-1}
\end{equation}
holds. For general $\hat{H}_{\mathrm{eff}}$ with complex eigenvalues, finding solutions can require imposing a maximum time $t < \tau_{\max}$ and a time-dependent $\hat{H} (t)$ even for time-independent $\hat{H}_{\mathrm{eff}}$. The imposition of $\tau_{\mathrm{max}}$ was already mentioned following Eq.~\eqref{eq:exacteffhamiltonian}. We now explain this finite-time requirement physically based on the post-selection interpretation.

First, note that the state in Eq.~\eqref{eq:initialstateancilla} generally does not factorize and therefore cannot directly serve as the product input of the Kraus construction in Eq.~\eqref{eq:krausoperator} when evolution is taken to start at $t=0$. This can be circumvented by initializing the composite system at an earlier time $t=-\epsilon$ as 
\begin{equation} \label{eq:initialization}
\ket{\Psi_{SE}(-\epsilon)} = \mathcal{N} \hat{F}(0)^{\frac{1}{2}} \ket{\Psi_S^u(0)} \otimes \ket{0_E},
\end{equation}
with the initial condition $\hat{F}(0)$ from Eq.~\eqref{eq:defops}. One then applies a unitary time evolution $\hat{U}_{\text{prep}}(\epsilon)$ that evolves this product state into the state of Eq.~\eqref{eq:initialstateancilla}. In Appendix~\ref{app:initialization} we show that the form of Eq.~\eqref{eq:initialization} allows for a unitary preparation of the state in Eq.~\eqref{eq:initialstateancilla} which is independent of $\ket{\Psi_S^u(0)}$.

Kraus operators are consequently defined as 
\begin{equation} \label{eq:krausoperatorancilla}
\begin{aligned}
\hat{\tilde{M}}_n(t) &\coloneqq \bra{n_E} \hat{U}(t) \hat{U}_{\mathrm{prep}}(\epsilon) \ket{0_E},\\
\hat{U}(t) &= \mathcal{T} e^{- i \int_0^{t} \! dt' \, \hat{H} (t')}, \\
\hat{\tilde{M}}_0(t) \ket{\Psi_S(-\epsilon)} &=\mathcal{N} \hat U_{\mathrm{eff}}^u(t) \ket{\Psi_S^u(0)},
\end{aligned}
\end{equation}
with the understanding that the post-selection is always performed after $t=0$. The tilde distinguishes this Kraus operator from our earlier definition in Eq.~\eqref{eq:krausoperator}, so that the effective time evolution operator has the same form as Eq.~\eqref{eq:timeevooperator}. The exact generator in Eq.~\eqref{eq:exacteffhamiltonian} is then defined using $\hat{\tilde{M}}_0(t)$ in place of $\hat{M}_0(t)$ for $t\geq0$. Such a unitary preparation procedure is standard in the quantum-information framework for quantum operations on qubits \cite{nielsen}, but the initialization of a many-body system in the correct state poses a further experimental challenge in addition to the fine-tuned construction of the Hamiltonian.

The need for an initial state with weight in both sectors becomes obvious in light of the Kraus map formalism: For active exponential gain to be possible, the post-selection probability has to increase. We recall the distinction between exponential and transient gain introduced in the discussion of Eq.~\eqref{eq:probsuperp}, as well as between active and passive gain, see Eq.~\eqref{eq:passivegain}. If $p_0(t)$ is to increase exponentially, its initial value cannot be unity. Thus, there has to be probability in the $\ket{1_E}$ sector at $t=0$. One might suspect that preparing the product state $\ket{\Psi_S(0)} \otimes \ket{1_E}$ and post-selecting for the $\ket{0_E}$ sector would also allow for an increase of $p_0(t)$ because $p_0(0)=0$. However, as argued before Eq.~\eqref{eq:timeevooperator}, the initial condition of the effective time evolution operator is only met if post-selecting for the initial environment state. Therefore, there needs to be nonzero initial probability in both sectors for an effective Hamiltonian evolution which allows for active exponential gain.

After post-selecting for the $\ket{0_E}$ sector at time $t$, the effective time evolution of the physical system state becomes 
\begin{equation} \label{eq:timeevogeneral}
\ket{\Psi_S (t)} = \mathcal{N} \frac{\hat U_{\mathrm{eff}}^u(t) \ket{\Psi^u_S(0)}}{\sqrt{p_0(t)}} = \mathcal{N} \frac{\mathcal{T} e^{-i \int_0^t \! dt' \, \hat{H}_{\mathrm{eff}}(t')}\ket{\Psi^u_S(0)}}{\sqrt{p_0 (t)}} \quad (t>0),
\end{equation}
with
\begin{equation} \label{eq:probgeneral}
p_0(t) = \mathcal{N}^2 \ev{\big[\hat U_{\mathrm{eff}}^u(t)\big]^{\dagger} \hat U_{\mathrm{eff}}^u(t)}{\Psi^u_S(0)} = \frac{1}{\ev{\underbrace{\hat{1} + \hat{g}^\dagger(t) \hat{g}(t)}_{\hat{F}(t)}}{\Psi_S(t)}}.
\end{equation}
In the last equation we used the relation 
\begin{equation} \label{eq:normalizationF}
1 = \mathcal{N}^2 \bra{\Psi_S^u(t)} \hat{F}(t) \ket{\Psi_S^u(t)}
\end{equation}
obtained from the normalization of the composite state (see Eqs.~\eqref{eq:initialstateancilla} and \eqref{eq:ancillaansatz}) as well as the definition of $\hat{F}(t)$ in Eq.~\eqref{eq:defops}. We emphasize that the initial probability $p_0 (0) \neq 1$ is not an arbitrary parameter. The boundary condition on $\hat{F}(0)$ must be carefully chosen to ensure that enough weight is in the $\ket{1_E}$ sector. This will be explained below.

The meaning of the operator $\hat{F}(t)$ is now apparent. From Eq.~\eqref{eq:ancillaresults2}, one derives that $\hat{F}^{-1}(t)$ fulfills the generalized von Neumann equation~\eqref{eq:generalizedvonNeumann} with the Hamiltonian $\hat{H}_{\mathrm{eff}}(t)$, which does not conserve probability. Therefore, $\hat{F}(t)$ is an inverse measure of relative probability for a fictitious ensemble residing in the system. The denominator of Eq.~\eqref{eq:probgeneral} is the expectation value of that measure, and its inverse, in return, yields the post-selection probability. Conceptually, one can view the environment as a virtual probability reservoir: the more probability is funneled into this reservoir, the larger the expectation value of $\hat{F}(t)$ becomes, and consequently, the less likely one is to find the particles in the system.

In the following, we assume a time-independent non-Hermitian Hamiltonian $\hat{H}_{\mathrm{eff}}$ to interpret the occurrence of eigenvalues with positive imaginary parts (active exponential gain). Inserting a right eigenstate $\ket*{\Psi^R}$ of $\hat{H}_{\mathrm{eff}}$ with eigenvalue $E + i \Gamma$ into Eq.~\eqref{eq:probgeneral}, one finds that the post-selection probability is given by
\begin{equation} \label{eq:gainprobability}
\ket*{\Psi^R}: \quad p_0(t) = e^{2\Gamma t} p_0(0).
\end{equation}
The imaginary part of an eigenvalue therefore quantifies the exponential increase or decrease of the post-selection probability of an eigenstate relative to its initial value.

It is clear that in the case $\Gamma>0$, once the post-selection probability reaches unity, i.e.,
\begin{equation} \label{eq:postselectionfailure}
 p_0(t^*) = 1 \Rightarrow t^* = -\frac{1}{2\Gamma}\ln p_0(0), 
\end{equation}
the description in terms of effective Hamiltonian dynamics must break down. A simple way to mathematically increase $t^*$ is to scale the boundary condition $\hat{F}(0)$ up by a constant, thereby decreasing $p_0(0)$ (cf.\ Eq.~\eqref{eq:probgeneral}).

Likewise, for an effective Hamiltonian with eigenvalues $\{ E_n + i \Gamma_n \}$ with at least one $\Gamma_n>0$, the time $t^*$ can be approximated by
\begin{equation}\label{eq:roughbound}
t^* \approx -\frac{1}{2 \Gamma_{\max}} \ln p_0(0)
\end{equation}
with $\Gamma_{\max} = \max\limits_{n} \{ \Gamma_n \}$. The approximation is very rough because it does not take into account the coherent fluctuations of the transient gain and loss that can significantly alter the behavior of $p_0(t)$, see Eq.~\eqref{eq:probsuperp}.

From Eq.~\eqref{eq:probgeneral}, one finds that the probabilistic interpretation of $p_0(t)$ is not violated as long as 
\begin{equation} \label{eq:initialconditiongeneral}
\ev{\big[\hat U_{\mathrm{eff}}^u(t)\big]^{\dagger} \hat U_{\mathrm{eff}}^u(t) - \hat{F}(0)}{\Psi_S(0)} \leq 0
\end{equation}
holds for all normalized initial states $\ket{\Psi_S(0)}$. In other words, the Hermitian operator $\big[\hat U_{\mathrm{eff}}^u(t)\big]^{\dagger} \hat U_{\mathrm{eff}}^u(t) - \hat{F}(0) \eqcolon \hat{\Lambda}(t)$ needs to be negative semi-definite. This condition also holds for time-dependent effective Hamiltonians. The exact, state-independent definition of $t^*$ is thus the first time for which one of the eigenvalues of $\hat{\Lambda}(t^*)$ becomes positive, and depends on the initial condition imposed on $\hat{F}(t)$.

The time $t^*$ need not be equal to the cutoff time $\tau_{\max}$ imposed on the solutions of Eqs.~\eqref{eq:ancillaresults} and~\eqref{eq:ancillaresults2}. It is the hard cutoff imposed by the laws of probability, and therefore an upper bound to $\tau_{\max}$. 

In order to quantify more precise bounds for $t^*$ than in Eq.~\eqref{eq:roughbound}, we assume that $\hat{H}_{\mathrm{eff}}$ is diagonalizable and consider a general linear combination of right eigenvectors $\ket{\Psi_S^u(0)} = \sum_n a_n \ket*{\Psi^R_n}$ with $a_n \in \mathbb{C}$. Then, an upper bound $B(t)$ of the post-selection probability is determined as
\begin{align}
p_0(t) &= \frac{\ev{e^{i \hat{H}_{\mathrm{eff}}^{\dagger}t} e^{-i\hat{H}_{\mathrm{eff}} t}}{\Psi_S(0)}}{\ev{\hat{F}(0)}{\Psi_S(0)}} \label{eq:explicitFdependence} \\ 
&= \mathcal{N}^2 \ev{e^{i \hat{H}_{\mathrm{eff}}^{\dagger}t} e^{-i\hat{H}_{\mathrm{eff}} t}}{\Psi_S^u(0)} \notag \\
&= \mathcal{N}^2 \left| \sum_{n,m} a_n a_m^* e^{i(E_m - E_n)t} e^{(\Gamma_m + \Gamma_n) t} \braket*{\Psi^R_m}{\Psi^R_n}\right| \notag \\
&\leq \mathcal{N}^2 \sum_{n, m} |a_n| |a_m| | \! \braket*{\Psi^R_m}{\Psi^R_n} \!| e^{(\Gamma_m + \Gamma_n) t} \notag \\
&\leq \mathcal{N}^2 e^{2 \Gamma_{\max}t} \sum_{n, m} |a_n| |a_m| | \! \braket*{\Psi^R_m}{\Psi^R_n} \!| \eqcolon B(t).
 \label{eq:envelope}
\end{align}

Because the definition of $t^*$ following Eq.~\eqref{eq:initialconditiongeneral} is state-independent, we define for a given initial condition $\hat{F}(0)$ the quantity
\begin{equation} \label{eq:Bsup}
B(t) \leq B_{\sup} (t) \coloneqq \max \big\{\sup_{\Psi_S(0)} B(t), 1\big\},
\end{equation}
where the supremum runs over all normalized initial system kets $\ket{\Psi_S(0)}$ (note that $B(t)$ does not depend on the norm of $\ket{\Psi_S^u(0)}$, cf.\ Eq.~\eqref{eq:explicitFdependence}).

If $\hat{H}_{\mathrm{eff}}$ possesses only real eigenvalues ($\Gamma_n = 0$), then $p_0(t)$ is bounded by a constant $B_{\sup}(t) \equiv B_{\sup}$. There are only superpositions of harmonic oscillations which are maximal for complete constructive interference of all non-orthogonal eigenvectors. This transient active gain could cause a collapse of the construction if $p_0(t)$ ever exceeds $1$, because contrary to the previous subsections, $-i \hat{H}^a_{\mathrm{eff}}$ is not necessarily negative semi-definite (cf.\ Eq.~\eqref{eq:semidefinite}). However, by rescaling 
\begin{equation} \label{eq:rescaling}
\hat{F}(0) \rightarrow B_{\sup} \hat{F}(0)
\end{equation}
and thus decreasing the initial $\ket{0_E}$ sector probability according to Eq.~\eqref{eq:explicitFdependence}, one obtains $t^* \to \infty$. Thus, a suitable $\hat{H}$ can, in principle, be found to obtain Hamiltonian evolution with $\hat{H}_{\mathrm{eff}}$ for all times, which is corroborated by the explicit construction in Refs.~\cite{WuPTSymmetry, PTpaper}. In particular, this applies to all \PT-symmetric Hamiltonians in the \PT-unbroken phase. 

The same holds true if all $\Gamma_n \leq 0$ because $B_{\sup}(t) \leq B_{\sup}(0) \equiv B_{\sup}$. This represents a genuine extension of the class of possible effective Hamiltonians compared to the last subsections because the eigenvalues of $\hat{H}^a_{\mathrm{eff}}$ are not restricted to the negative imaginary axis, allowing for transient active gain. 

If there is active exponential gain, i.e., at least one $\Gamma_n>0$, rescaling
\begin{equation} \label{eq:rescaling2}
\hat{F}(0) \rightarrow K B_{\sup}(0) \hat{F}(0)
\end{equation}
with $K>1$ allows one to avoid the collapse due to transient gain. The long-time exponential increase $p_0(t) \sim e^{2 \Gamma_{\max} t}$ for initial states with nonvanishing weight of the corresponding eigenstate will, however, inevitably lead to the probability crossing unity. A lower bound for $t^*$ is then found after dividing Eq.~\eqref{eq:envelope} evaluated at $t^*$ by the rescaling constant $K B_{\sup}(0)$, yielding
\begin{equation} \label{eq:betterbound}
\frac{1}{2 \Gamma_{\max}} \ln K \leq t^*.
\end{equation}
In the limit $K \to \infty$ and thus by decreasing the initial probability $p_0(0)$, $t^*$ can theoretically be extended to an arbitrarily large, finite value. The logarithmic dependence makes it unfeasible to achieve high values for $t^*$ in practice, because an exponentially small initial post-selection probability would preclude experimental accessibility of the non-Hermitian system at small times.

In conclusion, the ancilla approach is the most fruitful to derive the emergence of non-Hermitian dynamics because it allows for transient and exponential active gain and does not restrict the class of constructable Hamiltonians. It requires a careful engineering of a system-environment Hamiltonian, as well as a delicate unitary evolution that places the system in the desired initial state of Eq.~\eqref{eq:initialstateancilla}. No approximations are required, and the dynamics are equivalent to those generated by the exact Eq.~\eqref{eq:instantaneousselfenergy}. The simple two-state environment makes the problem analytically tractable. 

\begin{figure*}[p]
    \centering
    \includegraphics[width=\textwidth]{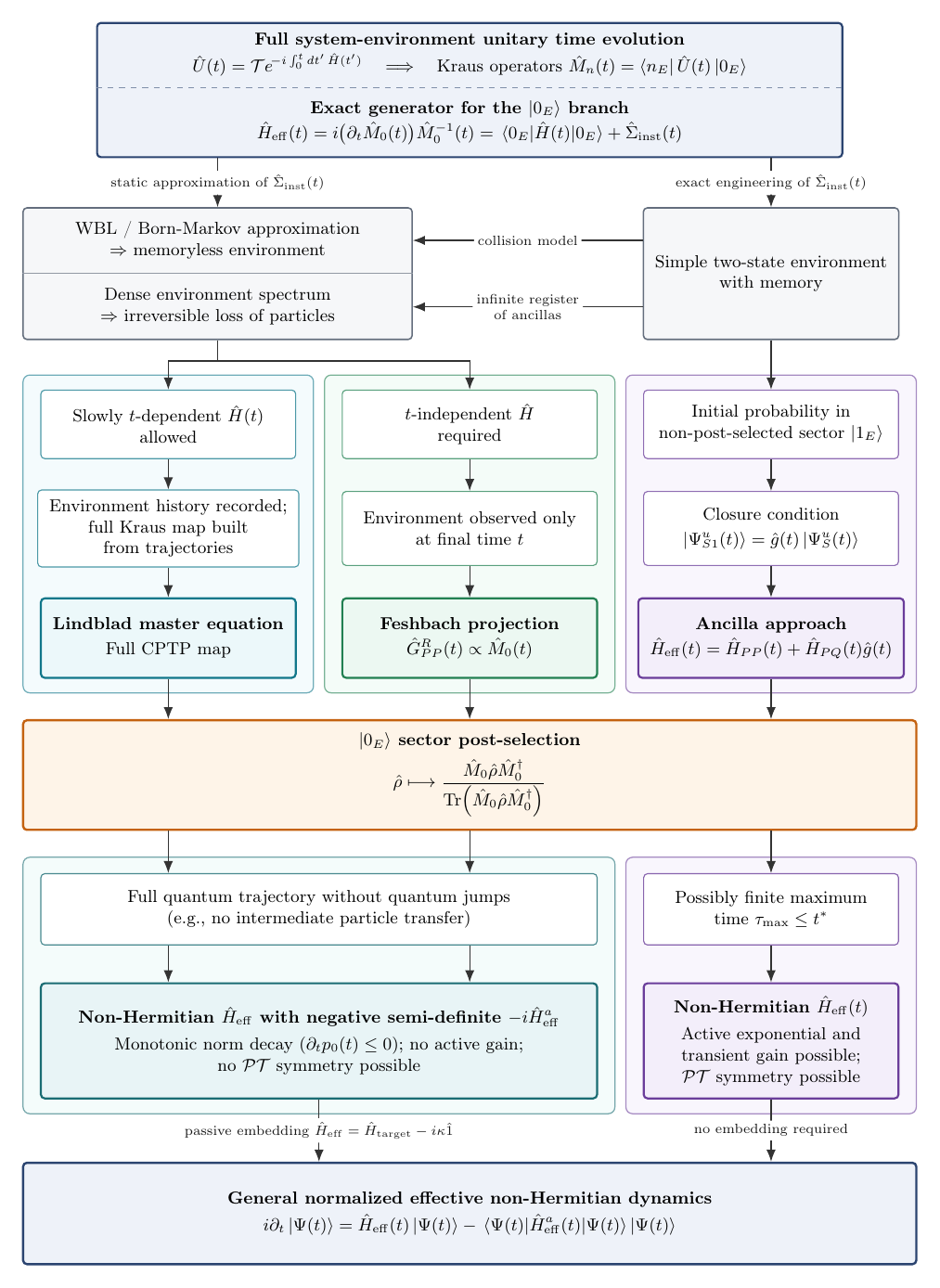}
    \caption{Three routes to effective non-Hermitian Hamiltonian dynamics. The columns distinguish Lindblad no-jump dynamics, Feshbach projection, and the ancilla approach.}
    \label{fig:nonhermitianflowchart}
\end{figure*}

All steps necessary in the three presented paths to non-Hermitian dynamics as well as the approximations involved and the restrictions on the effective Hamiltonians $\hat{H}_{\mathrm{eff}}$ are summarized in the flowchart of Fig.~\ref{fig:nonhermitianflowchart}.

\section{Interpretation of non-Hermitian dynamics}\label{sec:interpretation}
We have shown that each of the presented paths towards genuine non-Hermitian time evolution of quantum states involves a post-selection protocol and can be described by the Kraus map formalism. The first main consequence of this is that the states are normalized at all times, rendering the time evolution nonlinear. This stems from the projective nature of quantum mechanical measurements. 

A further consequence is that the composite Hilbert space $\mathcal{H}_S \otimes \mathcal{H}_E$ and therefore its inner product are fixed. This determines the inner products on the spaces $\mathcal{H}_S$ and $\mathcal{H}_E$ up to a rescaling constant, which is fixed once the normalization on one of the Hilbert spaces is specified. Therefore, a redefinition of the metric on $\mathcal{H}_S$ does not occur within this formalism. In particular, defining the biorthogonal inner product \cite{brody} in the case of an $\hat{\eta}$-pseudo-Hermitian Hamiltonian (see Eq.~\eqref{eq:pseudohermiticity}) as $\braket{\Psi_1}{\Psi_2}_{\hat{\eta}} \coloneqq \matrixel{\Psi_1}{\hat{\eta}}{\Psi_2}$ cannot be justified on physical grounds. 

In fact, from our discussion of the ancilla approach, we can exactly trace back where such a prescription might originate and why it is unphysical. For an $\hat{\eta}$-pseudo-Hermitian Hamiltonian $\hat{H}_{\mathrm{eff}}$, Eq.~\eqref{eq:ancillaresults2} with a time-independent $\hat{F}(t) \equiv \hat{F}$ implies that $\hat{F}$ is an $\hat{\eta}$-operator of the Hamiltonian. In that case, the inner product of two system-environment states $\ket{\Psi_{SE}(t)}$ and $\ket{\Phi_{SE}(t)}$ which initially are of the form of Eq.~\eqref{eq:initialstateancilla} and fulfill Eq.~\eqref{eq:ancillaansatz} is given by 
\begin{equation} \label{eq:etaconstruction}
\braket{\Phi_{SE}(t)}{\Psi_{SE}(t)} \propto \bra{\Phi_S^u(t)} \hat{\eta} \ket{\Psi_S^u(t)}.
\end{equation}
Importantly, this is the inner product in the full system-environment composite before post-selection, which evolves unitarily. Because genuine non-Hermitian dynamics exclusively describe the conditional sub-ensemble that survives post-selection onto the $\ket{0_E}$ branch, using this metric to evaluate physical observables on $\mathcal{H}_S$ improperly mixes in discarded environmental trajectories.

Consequently, the expectation value of an observable $\hat{O}$ in a right eigenstate $\ket*{\Psi^R}$ of $\hat{H}_{\mathrm{eff}}$ is evaluated as
\begin{equation} \label{eq:RRformalism}
\ev*{\hat{O}}_{\Psi^R} = \frac{\ev*{\hat{O}}{\Psi^R}}{\braket*{\Psi^R}}.
\end{equation}
This so-called ``RR'' formalism avoids the biorthonormal basis and retains the standard quantum mechanical definition. Furthermore, physical observables remain strictly Hermitian. This implies that the effective Hamiltonian acts solely as the generator of time evolution and cannot be regarded as an observable itself.

Despite the absence of left eigenvectors from Eq.~\eqref{eq:RRformalism}, the biorthonormalization condition 
\begin{equation} \label{eq:biorthonormalization}
\braket*{\Psi^L_n}{\Psi^R_m} = \delta_{nm}
\end{equation}
between left and right eigenstates is still useful as a mathematical tool. Determining both sets of eigenvectors is necessary to obtain the resolution of identity
\begin{equation} \label{eq:resolutionofidentity}
\hat{1} = \sum_n \ketbra*{\Psi_n^R}{\Psi_n^L},
\end{equation}
which is valid as long as $\hat{H}_{\mathrm{eff}}$ is diagonalizable \cite{brody}. This, in turn, allows for the spectral representation 
\begin{equation} \label{eq:spectralrepresentation}
\hat{H}_{\mathrm{eff}} = \sum_n (E_n + i \Gamma_n) \ketbra*{\Psi_n^R}{\Psi_n^L}
\end{equation}
of the effective Hamiltonian. Any expression containing only the operator $\hat{H}_{\mathrm{eff}}$ can likewise be expressed via a spectral representation. One example is the resolvent in Eq.~\eqref{eq:blockprojectedresolvent}, which is important for calculations in the frequency domain.

Furthermore, the biorthogonal basis can be used to derive nontrivial relations between physical quantities. For instance, establishing a generalized unitarity relation in non-Hermitian scattering theory \cite{BiorthogonalScattering} relies on the biorthogonality of left and right scattering states. 

\section{Non-Hermitian equilibrium ensembles} \label{sec:equilibrium}
As we have seen, establishing non-Hermitian dynamics requires the implementation of a post-selection protocol. Even for Hamiltonians with only real eigenvalues, the post-selection probability exhibits sinusoidal fluctuations for coherent superpositions of right eigenstates (see Eq.~\eqref{eq:probsuperp}). Therefore, it is natural to expect that the traditional notion of thermal equilibrium is generally not applicable, and that stationary states should instead be regarded as non-equilibrium steady states (NESS) fulfilling $\partial_t \hat{\rho}(t) = 0$. In the following, we will formalize this phenomenological insight precisely.

Thermal equilibrium requires the maximization of the von Neumann entropy $S = -\Tr (\hat{\rho} \ln \hat{\rho})$ given a number of constraints $\Tr \hat{\rho} \hat{O}_i = \overline{O}_i$ with specified observables $\{ \hat{O}_i \}$. In the Hermitian setting, choosing, for instance, the energy and particle number as constraints yields the grand canonical ensemble, describing a system in equilibrium with energy and particle reservoirs. For general observables $\{ \hat{O}_i \}$ and using a set of Lagrange multipliers $\{ \alpha_i \}$, the result is written as a Gibbs ensemble \cite{Jaynes, JaynesII}
\begin{equation} \label{eq:equilibriumstate}
\hat{\rho} = \frac{1}{Z} e^{-\sum_i \alpha_i \hat{O}_i},
\end{equation}
where the partition function $Z$ ensures normalization of the density operator. The logarithm of the partition function can be identified with the corresponding thermodynamic potential, bridging the gap between statistical mechanics and thermodynamics. 

Stationarity is a necessary condition for a thermal equilibrium ensemble. For a time-independent Hermitian generator of time evolution $\hat{H}$, choosing operators $\hat{O}_i$ that commute with $\hat{H}$ (in particular, $\hat{H}$ itself) ensures that the density operator in Eq.~\eqref{eq:equilibriumstate} is stationary. 

In the case of a non-Hermitian $\hat{H}_{\mathrm{eff}}$, stationarity implies that the density matrix before normalization fulfills $\hat{\rho}^u(t) = c(t) \hat{\rho}^u(0)$ with a real, positive function $c(t)$ (cf.\ Eq.~\eqref{eq:stationarityKraus}). Inserting this condition into the generalized von Neumann equation~\eqref{eq:generalizedvonNeumann} leads to
\begin{equation} \label{eq:demandstationarity}
\hat{H}_{\mathrm{eff}} \hat{\rho}^u(0) - \hat{\rho}^u(0) \hat{H}_{\mathrm{eff}}^{\dagger} = i \frac{\dot{c}(t)}{c(t)} \hat{\rho}^u(0) \eqcolon 2 i \Gamma \hat{\rho}^u(0),
\end{equation}
where we defined the real quantity $\dot{c}(t)/c(t) = 2 \Gamma$, which must be constant because $\hat{H}_{\mathrm{eff}}$ and $\hat{\rho}^u(0)$ are time-independent.

Away from exceptional points, $\hat{H}_{\mathrm{eff}}$ is diagonalizable with a complete, albeit generically non-orthonormal, set of right eigenvectors $\{ \ket*{\Psi^R_i} \}$, so we choose the ansatz $\hat{\rho} = \sum_{n,m} a_{nm} \ketbra*{\Psi^R_n}{\Psi^R_m}$ with $a_{nm} \in \mathbb{C}$ for a stationary and normalized density operator. Then, condition~\eqref{eq:demandstationarity} and positive semi-definiteness constrain $\hat{\rho}$ to the form
\begin{equation} \label{eq:statdensityop}
\hat{\rho} = \sum_{n,m} \pi_{nm} \delta_{E_n, E_m} \delta_{\Gamma_n, \Gamma} \delta_{\Gamma_m, \Gamma} \ketbra*{\Psi^R_n}{\Psi^R_m}.
\end{equation} 
It is composed solely of matrix elements between right eigenstates of $\hat{H}_{\mathrm{eff}}$ whose eigenvalues possess matching real parts and an imaginary part equal to the global constant $\Gamma$. This agrees with the observation made in Eq.~\eqref{eq:mixturesevo} that the decay rates of all right eigenstates in a mixture need to be identical if one demands stationarity\footnote{Without imposing positive semi-definiteness, the condition $\delta_{\Gamma_n,\Gamma}\delta_{\Gamma_m,\Gamma}$ would be replaced by the weaker condition $\delta_{\Gamma_n+\Gamma_m,2\Gamma}$. To see why positivity imposes the stronger constraint, note that stationarity forces the diagonal coefficient of any mode with $\Gamma_n\neq\Gamma$ to vanish. Positive semi-definiteness then forces all off-diagonal coefficients involving that mode to vanish as well, since $|a_{nm}|^2\leq a_{nn}a_{mm}$.}. The complex weights $\pi_{nm}$ have to be chosen so that $\hat{\rho}$ is normalized, Hermitian and positive semi-definite.

Equation~\eqref{eq:demandstationarity} can be rewritten as
\begin{equation} \label{eq:stationarityetaoperator}
	\hat{H}^{\prime}_{\mathrm{eff}} \hat{\rho}^u - \hat{\rho}^u \hat{H}^{\prime \dagger}_{\mathrm{eff}} = \hat{0},
\end{equation}
where we defined the shifted Hamiltonian $\hat{H}^{\prime}_{\mathrm{eff}} = \hat{H}_{\mathrm{eff}} -i\Gamma \hat{1}$. Equation~\eqref{eq:stationarityetaoperator} means that $\hat{\rho}^u$, if invertible, corresponds to $\hat{\eta}^{-1}$ of the $\hat{\eta}$-pseudo-Hermitian Hamiltonian $\hat{H}^{\prime}_{\mathrm{eff}}$ (cf.\ Eq.~\eqref{eq:pseudohermiticity}). The general theory of pseudo-Hermitian Hamiltonians \cite{Mosta_Part2} states that an invertible, positive definite $\hat{\eta}$-operator exists if and only if all eigenvalues of $\hat{H}^{\prime}_{\mathrm{eff}}$ are real, i.e., if and only if all eigenvalues of $\hat{H}_{\mathrm{eff}}$ have the same imaginary part $\Gamma$. This means that all diagonal terms in Eq.~\eqref{eq:statdensityop} can have nonzero weight and the whole Hilbert space can be spanned by the states in the mixture. It is physically intuitive that ergodicity would be violated if parts of the phase space were forbidden a priori. Because the imposed Gibbs state from Eq.~\eqref{eq:equilibriumstate} is invertible and positive definite, equilibrium in the form of a Gibbs ensemble cannot exist for any Hamiltonian possessing eigenvalues with different imaginary parts.

Provided that $\hat{H}_{\mathrm{eff}}$ possesses eigenvalues with a global imaginary part, one can set $\hat{\rho} \propto \hat{\eta}^{-1}$ and demand that it takes the form of the Gibbs ensemble in Eq.~\eqref{eq:equilibriumstate}. The specified observables and the Lagrange multipliers then have to fulfill
\begin{equation} \label{eq:constraintops}
\sum_i \alpha_i \hat{O}_i = \ln \hat{\eta} 
\end{equation}
up to an additive constant which can be absorbed into the normalization of $\hat{\rho}$. We have thus shown that, in contrast to the Hermitian situation, one cannot choose constraints on physical grounds by demanding that, e.g., the mean energy or particle number in the ensemble be a given value. This would be necessary to recover the traditional notion of equilibrium. Instead, for a given Hamiltonian $\hat{H}_{\mathrm{eff}}$ with a positive definite $\hat{\eta}$ fulfilling Eq.~\eqref{eq:stationarityetaoperator}, the operators $\hat{O}_i$ and Lagrange multipliers $\alpha_i$ are, themselves, constrained by Eq.~\eqref{eq:constraintops}. For example, assuming that the Hermitian part $\hat{H}^h_{\mathrm{eff}}$ is the appropriate energy observable of the system, the problem of finding constants $\alpha_1$ and $\alpha_2$ such that $\alpha_1 \hat{H}^h_{\mathrm{eff}} + \alpha_2 \hat{N} = \ln \hat{\eta}$ will not generally possess a solution, even with $\hat{\eta}$ not being unique \cite{Mosta_Part1}.

This concludes the reasoning that maximizing the entropy given a set of physical constraints does not lead to a stationary equilibrium ensemble in general non-Hermitian systems. In particular, a universal, form-invariant prescription for a thermal state which is automatically stationary is not provided by Eq.~\eqref{eq:equilibriumstate}. It is simply not the case that a full system-environment equilibrium can be described within the system by the non-Hermitian Hamiltonian. The post-selection process creates physically distinct situations. In particular, populating the right eigenstate of the non-Hermitian Hamiltonian belonging to the eigenvalue with the lowest real part is not a physical ground state and it is not expected to recreate the results computed, e.g., in the full Green's function framework without post-selection (see the results in Ref.~\cite{Lado} for a direct comparison).

While the notion of equilibrium is not applicable in the usual sense, a NESS in the form of Eq.~\eqref{eq:statdensityop} can in principle be constructed, which allows for the extraction of stationary properties of non-Hermitian systems.

\section{Conclusion} \label{sec:conclusion}
In this work, we have established a unified Kraus-map framework for three common approaches to non-Hermitian quantum dynamics: Feshbach projection, Lindblad no-jump evolution, and the ancilla approach. We have shown that all three can be understood in terms of post-selection, thereby clarifying their common origin as well as their distinct physical assumptions and limitations. While the exact conditional dynamics are generated by the time-local effective Hamiltonian including the instantaneous self-energy, the approaches build on different physical settings to render the description of the dynamics tractable.

In particular, we have identified the restrictions on the effective Hamiltonians arising from the Feshbach and Lindblad constructions and shown how the ancilla approach extends the accessible dynamics to include transient and active exponential gain. We have further related these gain and loss phenomena to the evolution of the relative post-selection probability, including the transient oscillations arising from the non-orthogonality of right eigenstates and the finite-time limitations imposed by active exponential gain.

Within the proposed framework, we showed that a consistent physical interpretation of non-Hermitian dynamics includes the use of the RR formalism for the calculation of expectation values in contrast to the biorthogonal inner product. Moreover, it was demonstrated that the classical notion of thermodynamic equilibrium is generally incompatible with post-selected, non-Hermitian dynamics and that stationary NESS should not be conflated with thermal states of the system-environment composite. This bottom-up physical justification provides a common basis for assessing competing prescriptions in the literature and may help resolve ongoing disagreements concerning the physical interpretation of non-Hermitian dynamics.

\appendix
\numberwithin{equation}{section}

\section{Derivation of a state-independent initialization procedure} \label{app:initialization}
We derive the form of the initial product state in Eq.~\eqref{eq:initialization} that allows for a state-independent preparation unitary $\hat{U}_{\mathrm{prep}}$. Using the expansion
\begin{equation} \label{eq:prepansatz}
\hat{U}_{\mathrm{prep}} = \sum_{m,n =0}^1 \hat{u}_{mn} \otimes \ketbra{m_E}{n_E},
\end{equation}
we conclude by comparing Eqs.~\eqref{eq:initialstateancilla} and \eqref{eq:ancillaansatz} with the initial state $\ket{\Psi_{SE}(-\epsilon)} = \ket{\Psi_{S}(-\epsilon)} \otimes \ket{0_E}$ that
\begin{equation} \label{eq:uconditions}
\begin{aligned}
\hat{u}_{00} \ket{\Psi_{S}(-\epsilon)} &= \mathcal{N} \ket{\Psi_S^u(0)},\\
\hat{u}_{10} \ket{\Psi_{S}(-\epsilon)} &= \mathcal{N} \hat{g}(0) \ket{\Psi_S^u(0)}.
\end{aligned}
\end{equation} 
Because this has to hold independently of the system state $\ket{\Psi_S^u(0)}$, we find that $\hat{u}_{10} = \hat{g}(0) \hat{u}_{00}$. Furthermore, the unitarity of $\hat{U}_{\mathrm{prep}}$ implies that
\begin{equation} \label{eq:unitarityimplication}
\hat{u}_{00}^\dagger \hat{u}_{00} + \hat{u}_{10}^\dagger  \hat{u}_{10} = \hat{u}_{00}^\dagger [\underbrace{\hat{1} + \hat{g}^\dagger(0) \hat{g}(0)}_{\hat{F}(0)}] \hat{u}_{00} \overset{!}{=} \hat{1},
\end{equation}
where we have identified $\hat{F}(0)$ from Eq.~\eqref{eq:defops}. With the choice $\hat{u}_{00} = \hat{F}(0)^{-\frac{1}{2}}$, this yields that
\begin{equation} \label{eq:statepreparationresult}
\ket{\Psi_{SE}(-\epsilon)} = \mathcal{N} \hat{F}(0)^{\frac{1}{2}} \ket{\Psi_S^u(0)} \otimes \ket{0_E}
\end{equation}
is a possible product state initialization that allows for a state-independent unitary $\hat{U}_{\mathrm{prep}}$ resulting in the desired $t=0$ state of Eq.~\eqref{eq:initialstateancilla}. A solution for the remaining two blocks of $\hat{U}_{\mathrm{prep}}$ is given by
\begin{equation} \label{eq:urest}
\hat{u}_{01} = -\hat{g}^\dagger(0) \hat{u}_{11} \quad \text{and} \quad \hat{u}_{11} = [\hat{1} + \hat{g}(0) \hat{g}^\dagger(0)]^{-\frac{1}{2}},
\end{equation}
which can be readily verified.

\bibliographystyle{unsrtnat}
\bibliography{references}

\end{document}